\documentclass[twocolumn]{aastex631}

\newcommand{\teff}{$T_{\mathrm{eff}}$}
\newcommand{\logg}{$\log g$}
\newcommand{\rearth}{$\mathrm{R_\oplus}$}
\newcommand{\fearth}{$\mathrm{F_\oplus}$}
\newcommand{\kep}{{\it Kepler}}
\newcommand{\ktwo}{{\it K2}}
\newcommand{\tess}{{\it TESS}}
\newcommand{\gaia}{{\it Gaia}}
\newcommand{\rprstar}{$R_{\mathrm{p}}/R_\star$}

\newcommand{\gbp}{$G_{\mathrm{BP}}$}
\newcommand{\grp}{$G_{\mathrm{RP}}$}
\newcommand{\msun}{$\mathrm{M}_\odot$}

\newcommand{\nstars}{\mbox{7993}}
\newcommand{\nplanets}{\mbox{9324}}

\usepackage{hyperref}
\usepackage{amsmath}

\makeatletter
\newcommand{\labeltext}[2]{%
  \@bsphack
  \csname phantomsection\endcsname 
  \def\@currentlabel{#1}{\label{#2}}%
  \@esphack
}

\shorttitle{The Gaia-Kepler-TESS-Host Stellar Properties Catalog}
\shortauthors{Berger et al.}

\graphicspath{{./}{figures/}}

\begin{document}

\title{The Gaia-Kepler-TESS-Host Stellar Properties Catalog:  Uniform Physical Parameters for 7993 Host Stars and 9324 Planets}

\correspondingauthor{Travis Berger}
\email{taberger@hawaii.edu}

\author[0000-0002-2580-3614]{Travis A. Berger}
\altaffiliation{NASA Postdoctoral Program Fellow}
\affiliation{Exoplanets and Stellar Astrophysics Laboratory, Code 667, NASA Goddard Space Flight Center, Greenbelt, MD, 20771, USA}

\author[0000-0001-5347-7062]{Joshua E. Schlieder}
\affiliation{Exoplanets and Stellar Astrophysics Laboratory, Code 667, NASA Goddard Space Flight Center, Greenbelt, MD, 20771, USA}

\author[0000-0001-8832-4488]{Daniel Huber}
\affiliation{Institute for Astronomy, University of Hawai`i, 2680 Woodlawn Drive, Honolulu, HI 96822, USA}

\begin{abstract}
We present the first homogeneous catalog of \kep, \ktwo, and \tess\ host stars and the corresponding catalog of exoplanet properties, which contain 7993 stars and 9324 planets, respectively. We used isochrone fitting and \gaia\ DR3 photometry, parallaxes, and spectrophotometric metallicities to compute precise, homogeneous \teff, \logg, masses, radii, mean stellar densities, luminosities, ages, distances, and V-band extinctions for 3248, 565, and 4180 \kep, \ktwo, and \tess\ stars, respectively. We compared our stellar properties to studies using fundamental and precise constraints, such as interferometry and asteroseismology, and find residual scatters of 2.8\%, 5.6\%, 5.0\%, and 31\%, with offsets of 0.2\%, 1.0\%, 1.2\%, and 0.7\% between our \teff, radii, masses, and ages and those in the literature, respectively. In addition, we compute planet radii, semimajor axes, and incident fluxes for 4281, 676, and 4367 \kep, \ktwo, and \tess\ planets, respectively, and find that the exoplanet radius gap is less prominent in the \ktwo, \tess, and combined samples than it is in the \kep\ sample alone. We suspect this difference is largely due to heterogeneous planet-to-star radius ratios, shorter time baselines of \ktwo\ and \tess, and smaller sample sizes. Finally, we identify a clear radius inflation trend in our large sample of hot Jupiters and find 150 hot sub-Neptunian desert planets, in addition to a population of over 400 young host stars as potential opportunities for testing theories of planet formation and evolution.
\end{abstract}

\section{Introduction}

The \kep, \ktwo, and \tess\ Missions have enabled the detection of thousands of transiting exoplanets, from close-in hot Jupiters to Earth-analogs, transforming our understanding of exoplanet demographics and planet formation and evolution. Until now, most analyses have focused either on the detection and characterization of individual systems \citep[e.g.][]{holman10,howell12,hebrard13,ciceri15,vanderburg15,schlieder16,Cloutier2020} or the characterization of populations of exoplanets from a particular mission \citep{Dressing2015,Petigura2017,Johnson2017,Fulton2017,Fulton2018,Berger2018c,Stassun2019,Berger2020a,Berger2020b,Hardegree2020,Zink2021}, but none have performed a homogeneous characterization of the entire population of host stars to transiting exoplanets.

Homogeneous determinations of stellar properties are important because of the potential for strong systematics between different observables methodologies converting those observables into fundamental physical parameters. For instance, it has been shown that the same spectrum analyzed by different pipelines produces different results \citep{Torres2012}, with discrepancies exceeding uncertainties in some cases \citep{Tayar2022}. While it is difficult to determine fundamental parameters accurately in an absolute sense, comparing stellar parameters homogeneously and then comparing them to each other remains an easier task. 

Here, we present the first homogeneous characterization of the population of \kep, \ktwo, and \tess\ transiting planet host stars using \texttt{isoclassify} \citep{Huber2017,Berger2020a}, only leveraging observables from \gaia\ DR3 \citep{gaia1,GaiaEDR3,GaiaDR3,Babusiaux2022}: parallaxes, \gbp\ and \grp\ photometry, spectrophotometric metallicities, and measured positions. We re-derive stellar \teff, \logg, radii, masses, densities, luminosities, and ages for \nstars\ \kep, \ktwo, and \tess\ hosts and compare our performance to current best estimates from interferometry, asteroseismology, and clusters. Finally, we re-compute planet radii, semi-major axes, and incident fluxes by combining our new stellar properties with results from planet transit-fitting pipelines and compare the host star and planet populations of \kep, \ktwo, and \tess.

\section{Methodology}
\label{sec:methods}

\subsection{Sample Selection}
\label{sec:sampsel}

For \kep\ hosts, we used the Cumulative KOI table at the NASA Exoplanet Archive \citep{Batalha2013,burke14,rowe15,mullally15,Coughlin2016,Thompson2018,kepcumulative}\footnote{accessed 11/5/21}, for \ktwo\ hosts, we utilized the \cite{Zink2021} catalog, and for \tess\ hosts, we used the \tess\ Project Candidates table \citep[e.g.][]{Guerrero2021} from the NASA Exoplanet Archive\footnote{accessed 7/27/22}. We followed a procedure identical to \cite{Berger2018c} to crossmatch \kep\ hosts to \gaia\ DR3 sources, but instead used TOPCAT \citep{topcat} to perform the sky crossmatch. For \ktwo\ and \tess\ hosts, we first selected all \gaia\ DR3 sources within 4'' and determined the minimum of the resulting angular distance histograms (1.5'' for \ktwo\ and 1.0'' for \tess). We then removed all matches with angular distances larger than that angular distance. For the remaining matches where one Ecliptic Plane Input Catalog \citep[EPIC,][]{huber16} or \tess\ Input Catalog \citep[TIC,][]{Stassun2019} star has multiple \gaia\ DR3 sources, we choose the smallest absolute $G-K_p$ and $G-T$ discrepancy relative to the median magnitude difference for a star with that $K_p$ or $T$ magnitude flag.

From there, we removed stars without positive \gaia\ parallaxes or \gbp/\grp\ photometry. We next removed all planet candidates already flagged as false positives in each planet sample. After this step, we retained 3309 \kep, 577 \ktwo, and 4198 \tess\ host stars.

\subsection{Input Parameters}\label{sec:inputs}

\begin{figure}
\resizebox{\hsize}{!}{\includegraphics{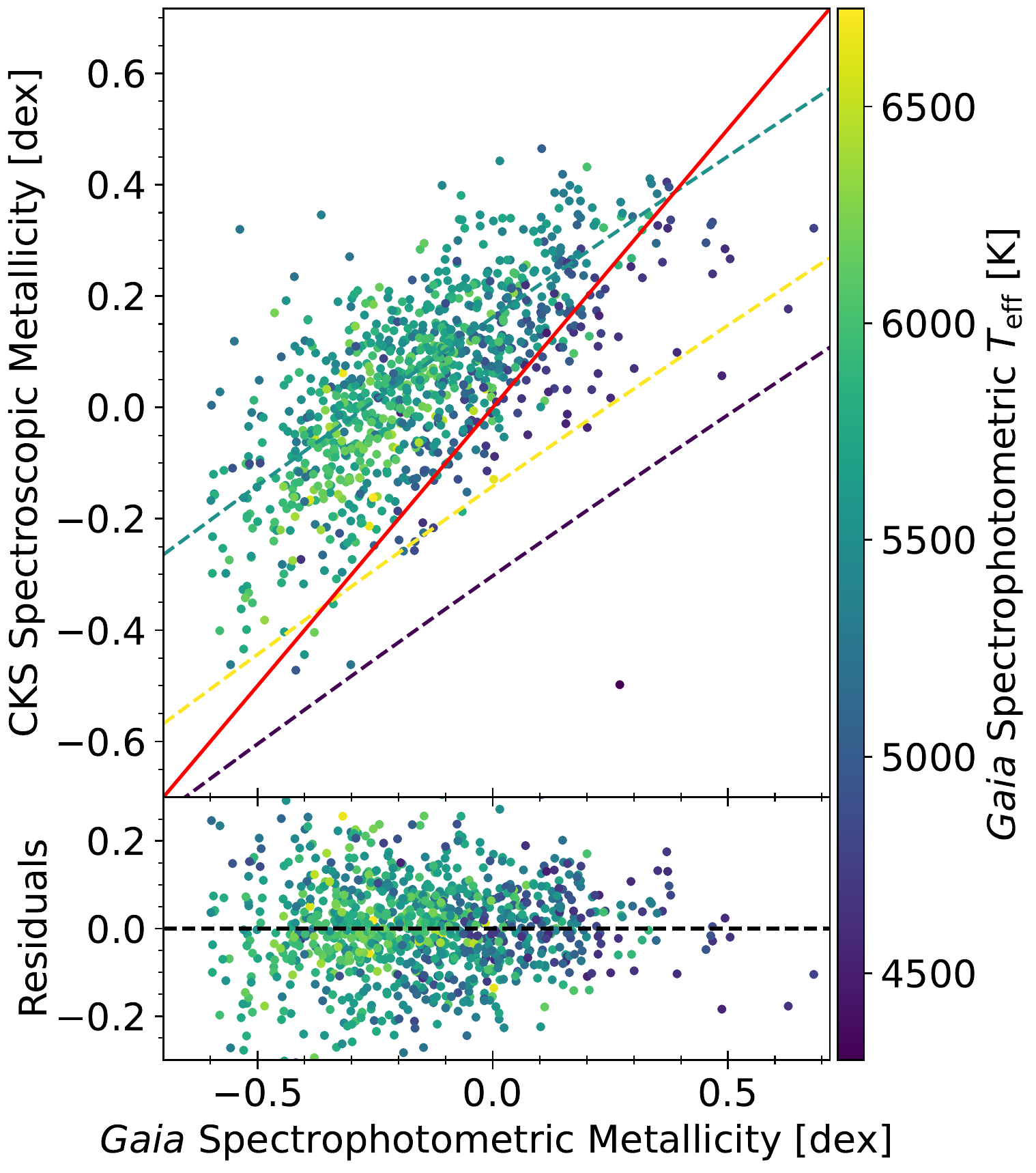}}
\caption{California-\kep\ Survey \citep[CKS,][]{Petigura2017} spectroscopic metallicities versus those from \gaia\ DR3 spectrophotometry \citep{GaiaDR3,Andrae2022,Creevey2022,Fouesneau2022}. In the top panel, we plot a direct comparison between the CKS and \gaia\ metallicities, where the solid red line represents agreement and the colored curves show the best-fit polynomial relations used to correct \gaia\ metallicities for stars of different \teff. The bottom panel shows the residuals between the \gaia\ and CKS data after applying the corrections to the \gaia\ metallicities and the black-dashed line represents agreement. Row 1 of Table \ref{tab:metfix} contains the polynomial relation used to correct \gaia\ metallicities.}\label{fig:metfix}
\end{figure}

\begin{deluxetable}{ll}
\tabletypesize{\scriptsize}
\tablecolumns{2}
\tablecaption{CKS-\gaia\ Spectrophotometric Metallicites}
\tablehead{\colhead{Bounds} & \colhead{Adopted Metallicity}}
\def\arraystretch{1.0}
\startdata
\begin{tabular}{@{}c@{}}1. 4000 $<$ $T_{\mathrm{eff},G}$ $<$ 7000 K \\ $\cap$ [M/H]$_G$ $>$ --2.0 dex \\ \end{tabular} & \begin{tabular}{@{}c@{}}[Fe/H]$_{\mathrm{adopt}}$ = --5.30 +  0.591 [M/H]$_G$ \\ -- 0.0257 [M/H]$_G^2$ +  0.00193 $T_{\mathrm{eff},G}$ \\ -- 0.000000171 $T_{\mathrm{eff},G}^2$\end{tabular}\\
2. Other & [Fe/H]$_{\mathrm{adopt}}$ = 0.16 + 0.605 [M/H]$_G$ \\
\enddata
\tablecomments{CKS-based polynomial relations used to fix \gaia\ spectrophotometric metallicities. The $G$ denotes \gaia\ spectrophotometric parameters, while [Fe/H]$_{\mathrm{adopt}}$ represent our adopted metallicities. We plot the results of these relations in Figure \ref{fig:metfix}.}
\vspace{-1.0cm}
\label{tab:metfix}
\end{deluxetable}

To create a homogeneous catalog for the combined \kep+\ktwo+\tess\ host star samples, we used \gaia\ DR3 \gbp\ and \grp\ as our photometric constraints. We chose these two passbands due to their mmag precision \citep{Riello2021}, color-\teff\ sensitivity, and \grp's relative insensitivity to extinction. The typical host star in our sample experiences $A_V$ = 0.15 mag, which we correct for with an extinction map. In addition, \gaia\ photometry has a $\lesssim$1'' angular resolution \citep{Bruijne2015} compared to 2MASS's 4'' \citep{skrutskie06}, which minimizes issues of photometric blending. We used \citet{Riello2021} equations C.1-C.3 to correct saturated photometry, where relevant. While we did not modify the formal, $input$ uncertainties on the \gaia\ photometry, their mmag precision makes them difficult to use with a relatively sparse pre-computed grid like \texttt{isoclassify} uses; hence, we will detail the required modifications to \texttt{isoclassify} in \S\ref{sec:modelgrid}.

In addition to the \gbp\ and \grp\ photometry, we utilized \gaia\ DR3 parallaxes \citep{Lindegren2021a} with zeropoint corrections \citep{Lindegren2021b} and J2000 positions. Where possible, we used spectrophotometric metallicities \citep[\texttt{mh\_gspphot},][]{Creevey2022,Fouesneau2022,Andrae2022}. Given that the spectrophotometric metallicities exhibit strong systematic errors \citep{Andrae2022}, we corrected them using California-\kep\ Survey \citep[CKS,][]{Petigura2017} metallicities by fitting a quadratic polynomial as a function of spectrophotometric \teff\ (\texttt{teff\_gspphot}) and metallicity to the CKS metallicities. We only applied this relation to stars with \gaia\ spectrophotometric \teff\ between 4000 and 7000 K and metallicities $>$--2.0 dex. For stars outside that range, we used a line representing the median offset between the \gaia\ spectrophotometric metallicities and the CKS. We display the CKS-\gaia\ metallicity comparison in Figure \ref{fig:metfix} and corresponding relations in Table \ref{tab:metfix}. These homogeneous metallicities are available for a larger fraction of host stars, unlike the heterogeneous metallicities of LAMOST \citep{LAMOSTDR5}, APOGEE \citep{APOGEEDR14}, and the CKS \citep{Petigura2017} used in \citet{Berger2020a}.

Table \ref{tab:input} shows the parameters that we input into \texttt{isoclassify}. The metallicity provenance column can be one of three values:  Poly, Med, or None. For stars with Poly provenances, we used the polynomial equation in row 1 of Table \ref{tab:metfix} and adopt an uncertainty of 0.15 dex as in \citet{Berger2020a} and for those with Med provenances (outside the \teff\ and metallicity range of CKS stars), we used the linear equation in row 2 of Table \ref{tab:metfix} and adopt an uncertainty of 0.20 dex. These uncertainties are conservative by design, as \citet{Furlan2018} shows that systematic metallicity uncertainties between spectroscopic pipelines are typically $\geq$0.1 dex. Likewise, the scatter in the residuals of Figure \ref{fig:metfix} is $\approx$0.11 dex. Stars with None provenances lack \gaia\ DR3 spectrophotometric metallicities, and for these stars \texttt{isoclassify} defaults to a broad ($\gtrsim$0.20 dex), Gaussian solar metallicity prior.

\begin{deluxetable*}{lccccccr}
\tabletypesize{\scriptsize}
\tablewidth{0pt}
\tablecolumns{8}
\tablecaption{\gaia-\kep-\tess-Host Stellar Input Parameters}
\tablehead{\colhead{Star ID} & \colhead{DR3 Source ID} & \colhead{\gbp\ [mag]} & \colhead{\grp\ [mag]} & \colhead{$\pi$ [mas]} & \colhead{[Fe/H]} & \colhead{[Fe/H] Prov} & \colhead{RUWE}}
\startdata
kic10858832 & 2129500383713365248 & 14.5940 $\pm$ 0.0030 & 13.7443 $\pm$ 0.0039 & 1.1711 $\pm$ 0.0135 & 0.200 $\pm$ 0.150 & Poly & 1.003 \\
kic2571238 & 2051106987063242880 & 12.2374 $\pm$ 0.0028 & 11.3370 $\pm$ 0.0038 & 4.5680 $\pm$ 0.0087 & 0.137 $\pm$ 0.150 & Poly & 0.824 \\
kic8628665 & 2126430409811576704 & 15.2767 $\pm$ 0.0031 & 14.4036 $\pm$ 0.0039 & 0.9280 $\pm$ 0.0185 & 0.072 $\pm$ 0.150 & Poly & 1.016 \\
kic10328393 & 2130683904901068800 & 14.4843 $\pm$ 0.0032 & 13.3366 $\pm$ 0.0039 & 2.8866 $\pm$ 0.0133 & -0.003 $\pm$ 0.150 & Poly & 1.010 \\
epic210577548 & 44062348664673152 & 13.6642 $\pm$ 0.0029 & 12.4527 $\pm$ 0.0038 & 3.5165 $\pm$ 0.0151 & -0.027 $\pm$ 0.150 & Poly & 0.939 \\
epic220294712 & 2538769088655260416 & 12.5499 $\pm$ 0.0028 & 11.8115 $\pm$ 0.0038 & 2.3343 $\pm$ 0.0151 & -0.092 $\pm$ 0.150 & Poly & 1.001 \\
epic211711685 & 611385506505666688 & 12.7334 $\pm$ 0.0028 & 11.8419 $\pm$ 0.0038 & 3.4644 $\pm$ 0.0134 & 0.188 $\pm$ 0.150 & Poly & 0.944 \\
epic251584580 & 3658535369882355712 & 14.8086 $\pm$ 0.0033 & 13.6446 $\pm$ 0.0041 & 2.6322 $\pm$ 0.0220 & -0.289 $\pm$ 0.150 & Poly & 1.053 \\
tic429501231 & 3423681228084838912 & 13.6640 $\pm$ 0.0032 & 12.7718 $\pm$ 0.0048 & 1.8895 $\pm$ 0.0285 & -0.204 $\pm$ 0.150 & Poly & 1.533 \\
tic255685030 & 5501542395359045504 & 11.4133 $\pm$ 0.0029 & 10.5290 $\pm$ 0.0038 & 6.3511 $\pm$ 0.0124 & 0.226 $\pm$ 0.150 & Poly & 0.990 \\
tic238624131 & 378265951674712448 & 12.4537 $\pm$ 0.0028 & 11.5585 $\pm$ 0.0038 & 3.0227 $\pm$ 0.0132 & 0.188 $\pm$ 0.150 & Poly & 1.043 \\
tic391903064 & 5212899427468919296 & 9.5947 $\pm$ 0.0075 & 8.7303 $\pm$ 0.0061 & 12.6898 $\pm$ 0.0105 & 0.154 $\pm$ 0.150 & Poly & 0.893 \\
\enddata
\tablecomments{Star ID (kic for \kep, epic for \ktwo, and tic for \tess), \gaia\ DR3 source ID, \gbp-mag, \grp-mag, parallax, metallicity, and metallicity provenance parameters and their uncertainties, which were provided as input to \texttt{isoclassify}. We also provide \gaia\ DR3 RUWE values for posterity. The [Fe/H] Prov column has three possible values:  Poly, Med, and None. Poly values denote our use of the CKS-based polynomial method to fix the \gaia\ spectrophotometric metallicity (row 1 of Table \ref{tab:metfix}), Med values denote our use of the median offset line (row 2 of Table \ref{fig:metfix}), and None values denote stars that lack \gaia\ spectrophotometric metallicities, where \texttt{isoclassify} defaults to a broad, Gaussian solar metallicity prior. A subset of our input parameters is provided here to illustrate the form and format. The full table, in machine-readable format, can be found online.}
\vspace{-0.5cm}
\label{tab:input}
\end{deluxetable*}

\subsection{Isochrone Fitting}\label{sec:modelgrid}

\begin{figure}
\resizebox{\hsize}{!}{\includegraphics{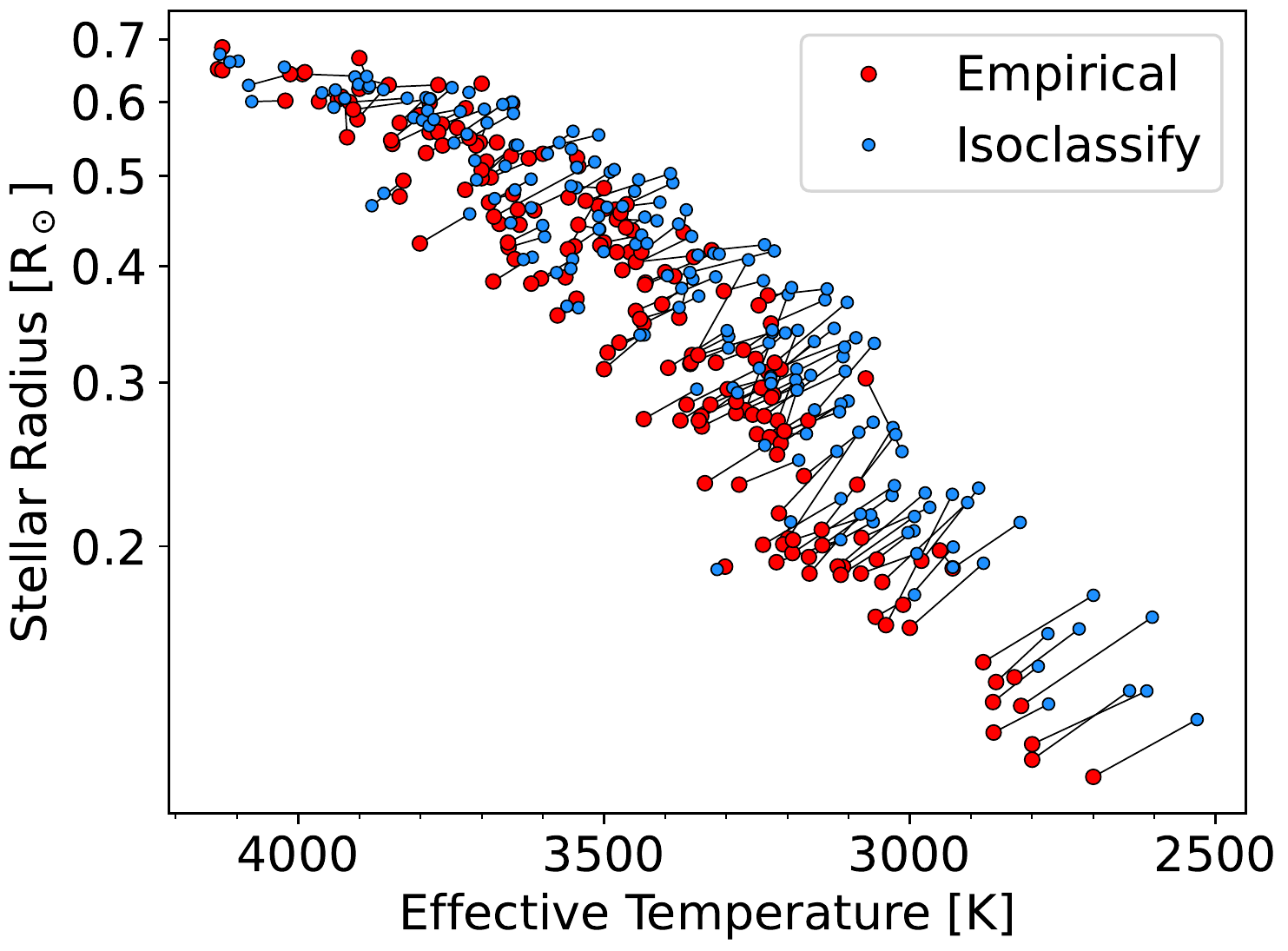}}
\caption{Stellar radius versus effective temperature for the \citet{mann15} M-dwarfs, where the red points represent the empirical data and the blue points are the \texttt{isoclassify} output using the plain PARSEC models \citep{bressan12,Chen2019}. We draw black lines connecting estimates for the same star. Because of this discrepancy, we use empirical corrections to constrain the properties of low mass stars.} 
\label{fig:mdwarfcomp}
\end{figure}

\begin{figure*}
\resizebox{0.5\hsize}{!}{\includegraphics{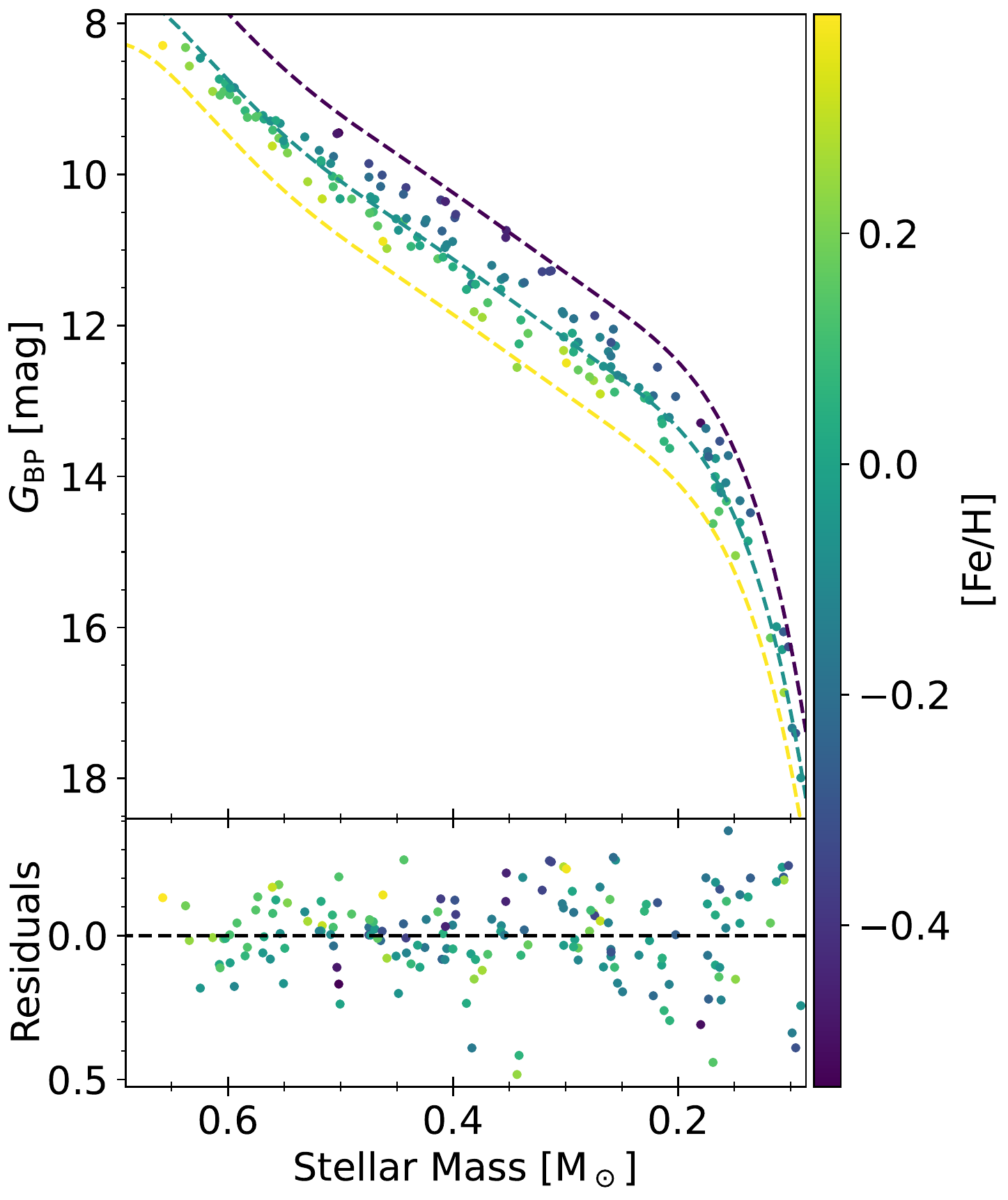}}
\resizebox{0.5\hsize}{!}{\includegraphics{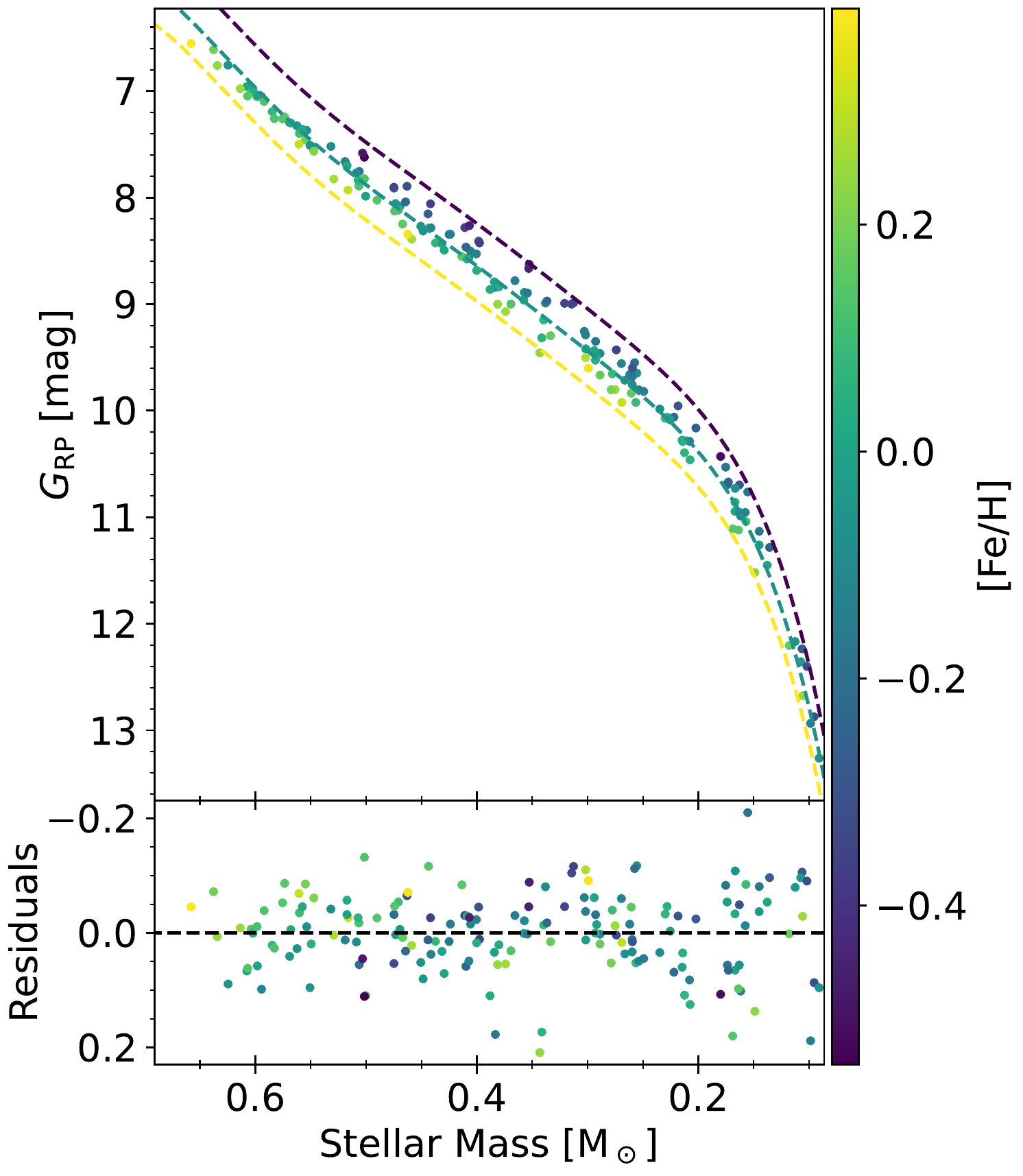}}
\resizebox{0.5\hsize}{!}{\includegraphics{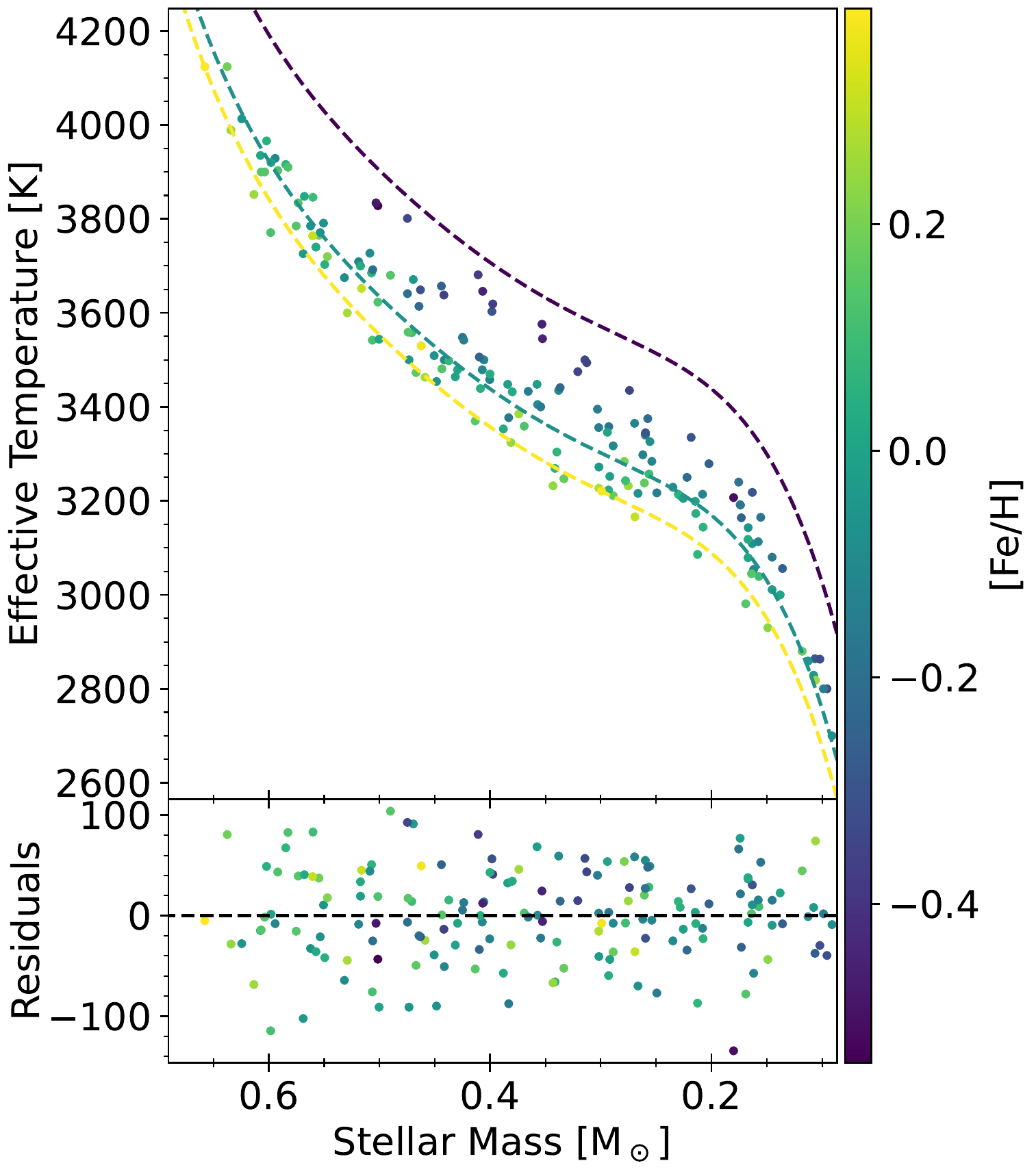}}
\resizebox{0.5\hsize}{!}{\includegraphics{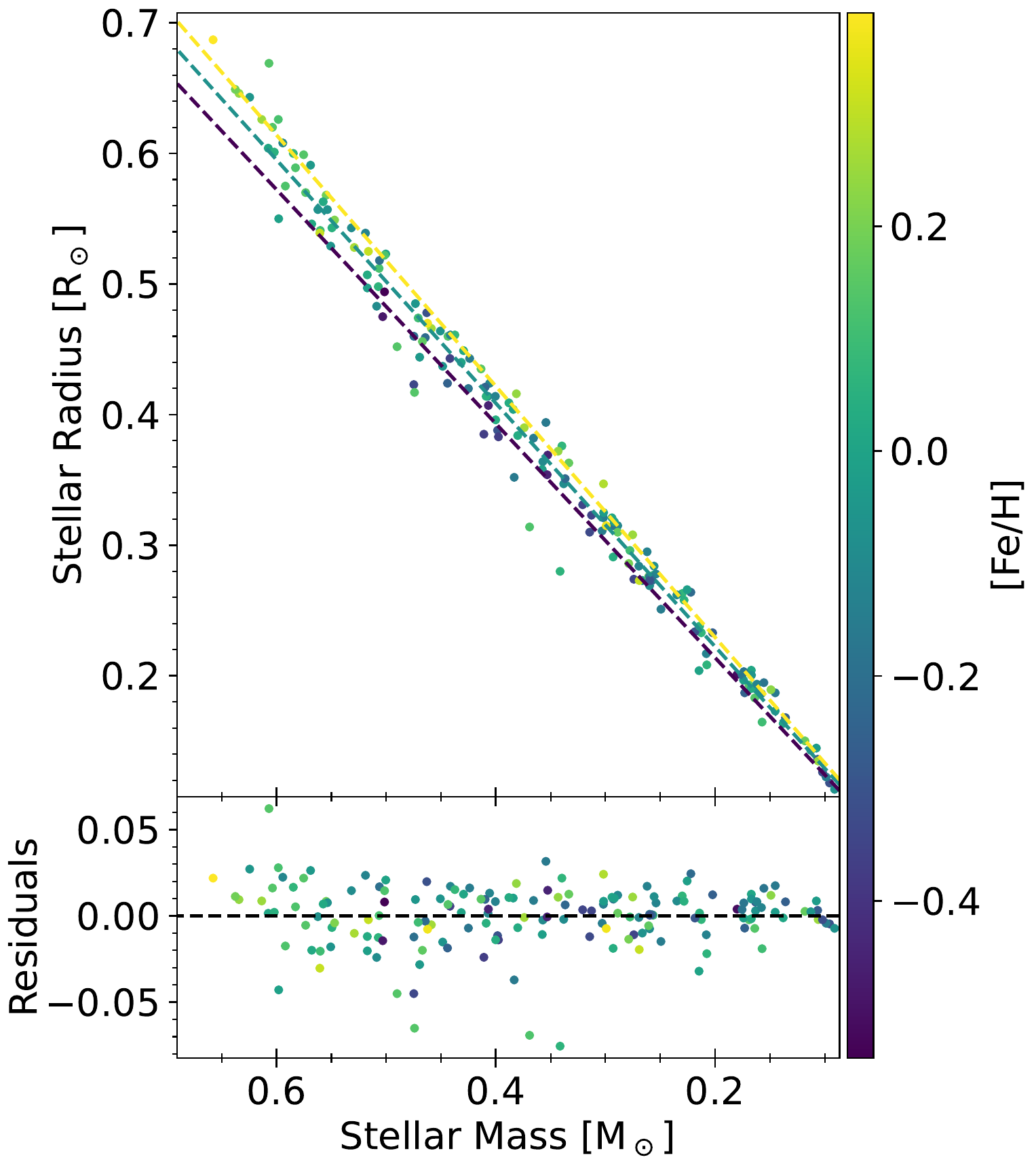}}
\caption{Empirical measurements (absolute \gbp, absolute \grp, \teff, radius) versus stellar mass colored by metallicity for the \cite{mann15} sample of M-dwarfs. We derived the masses using the \citet{Mann2019} \texttt{M\_-M\_K-} code. Yellow, green, and indigo dashed lines show the best-fit relations for [Fe/H] = --0.6, 0.0, and 0.5 dex, and residuals show the observed values subtracted by the predicted value of the polynomial with a corresponding mass and metallicity, where the black dashed line indicates agreement. We include the numerical relations in Table \ref{tab:emprels}.}
\label{fig:mdwarfemprels}
\end{figure*}

\begin{figure}
\resizebox{\hsize}{!}{\includegraphics{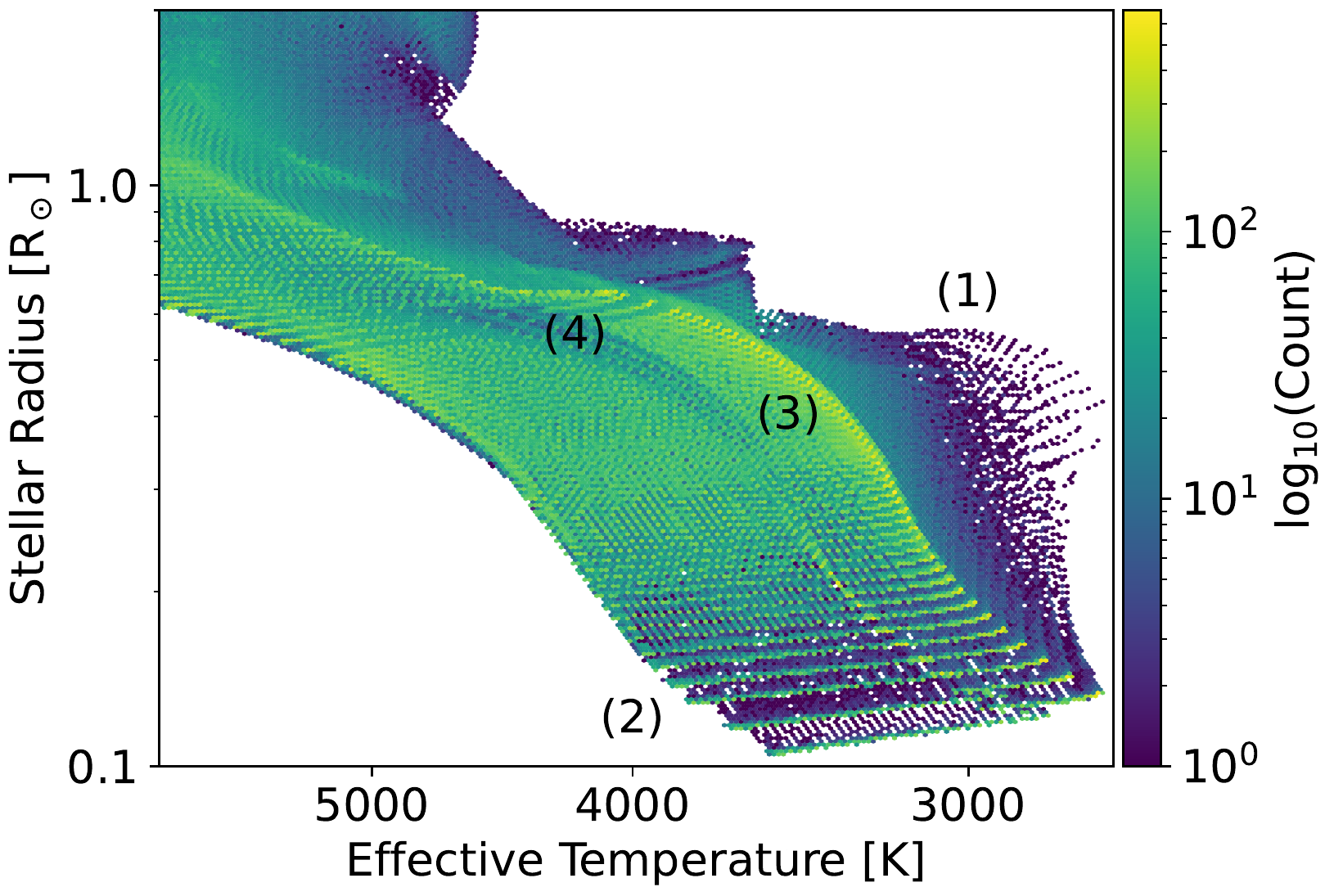}}
\caption{Stellar radius versus effective temperature for the empirically modified, interpolated PARSEC model grid utilized in this work. The logarithmic number density of models is represented by the color of each bin.} 
\label{fig:parsecemp}
\end{figure}

\begin{figure}
\resizebox{\hsize}{!}{\includegraphics{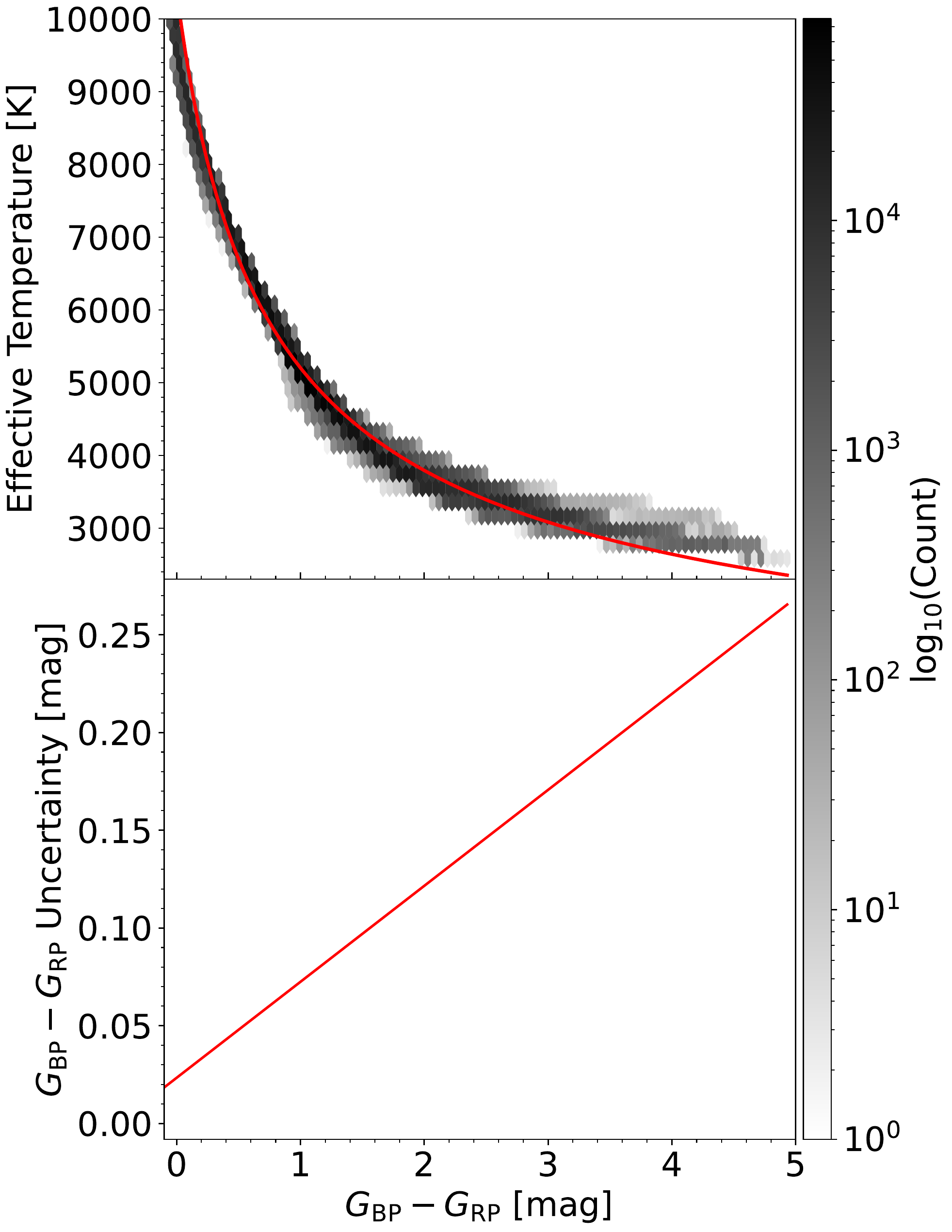}}
\caption{We place a floor on \gbp--\grp\ color uncertainties through our empirically modified PARSEC model grid. $Top$: \teff\ versus \gbp--\grp\ color for our model grid, with the logarithmic density of points illustrated by the two-dimensional greyscale histogram with the corresponding colorbar. We plot our best-fit \texttt{SmoothlyBrokenPowerLaw1D} in red. $Bottom$: The red line represents the required \gbp--\grp\ minimum uncertainty to reach a 3\% \teff\ error for all stars, dependent on their \gbp--\grp\ color.} 
\label{fig:tefferrfix}
\end{figure}

\begin{deluxetable*}{cl}
\tabletypesize{\scriptsize}
\tablecolumns{2}
\tablecaption{Polynomial Empirical Relations for M-dwarfs}
\tablehead{\colhead{Parameter} & \colhead{Polynomial Empirical Relation}}
\def\arraystretch{1.0}
\startdata
$G$ & 24.5898 -- 183.9893 $\mathrm{M}_\star$ + 1120.1562 $\mathrm{M}^2_\star$ -- 3770.5221 $\mathrm{M}^3_\star$ + 7016.0366 $\mathrm{M}^4_\star$ -- 6804.3093 $\mathrm{M}^5_\star$ + 2672.7206 $\mathrm{M}^6_\star$ + 0.8813 [Fe/H] \\
\gbp\ & 34.4834 -- 318.4227 $\mathrm{M}_\star$ + 2022.7769 $\mathrm{M}^2_\star$ -- 6932.4997 $\mathrm{M}^3_\star$ + 13040.8749 $\mathrm{M}^4_\star$ -- 12736.0413 $\mathrm{M}^5_\star$ + 5032.4396 $\mathrm{M}^6_\star$ + 1.4652 [Fe/H] \\
\grp\ & 22.1215 -- 165.2395 $\mathrm{M}_\star$ + 996.5293 $\mathrm{M}^2_\star$ -- 3330.1792 $\mathrm{M}^3_\star$ + 6149.4185 $\mathrm{M}^4_\star$ -- 5913.3984 $\mathrm{M}^5_\star$ + 2302.3600 $\mathrm{M}^6_\star$ + 0.6634 [Fe/H] \\
\teff\ & 1352.4644 + 22894.5349 $\mathrm{M}_\star$ -- 113739.3956 $\mathrm{M}^2_\star$ + 286679.2022 $\mathrm{M}^3_\star$ -- 350226.7290 $\mathrm{M}^4_\star$ + 170350.2318 $\mathrm{M}^5_\star$ -- 292.0425 [Fe/H] + 260.4901 [Fe/H]$^2$ \\
$R_\star$ & (0.0361 + 0.9316 $\mathrm{M}_\star$) $\times$ (1 + 0.0638 [Fe/H]) \\
\enddata
\tablecomments{Empirical relations for M-dwarfs based on stellar mass and metallicity. The PARSEC models are defined based on their masses and metallicities, so we compute empirical relations for the observables as a function of mass and metallicity. The \gaia\ photometric relations in this table are for absolute magnitudes, and the \teff\ and radius relations were used to amend the \logg\ and mean stellar densities of the models. These relations are only used for main sequence models with masses $<$ 0.63 \msun\ and metallicities $>$ --0.5 dex. We plot these relations in Figure \ref{fig:mdwarfemprels}. We also include \gaia\ $G$ for posterity.} \label{tab:emprels}
\end{deluxetable*}

In order to determine fundamental physical parameters from the input observables adopted from \gaia\ DR3, we used \texttt{isoclassify} \citep{Huber2017,Berger2020a} in combination with a custom-interpolated PARSEC evolutionary model grid \citep{bressan12}. We used \texttt{kiauhoku} \citep{Claytor2020} to finely interpolate the coarse models to 0.05 dex steps in metallicity between --2.0 to 0.5 dex, 117 masses between 0.1 and 5.0 \msun, and ages between 0.01 Gyr and 30 Gyr. We used 100 equivalent evolutionary points \citep[EEPs,][]{dotter16} for the pre-main sequence, 250 for the main sequence, 100 for the subgiant branch, and 200 for the red giant branch up to the tip. We used YBC bolometric corrections \citep{Chen2019} to compute synthetic photometry for our interpolated models.

Upon applying this grid to the sample of \citet{mann15,Mann2019} M-dwarfs crossmatched with \gaia\ photometry and parallaxes, we found large discrepancies between the measured physical parameters and those produced by our isochrone fitting analysis. Figure \ref{fig:mdwarfcomp} shows that large systematics are present, and that the plain PARSEC grid produces systematically larger and cooler parameters for M-dwarfs than empirically measured. Therefore, we used the \citet{Mann2019} \texttt{M\_-M\_K-} code, which uses 2MASS $K_s$ magnitudes and \gaia\ parallaxes to compute stellar masses, to compute the empirical masses of the \cite{mann15} sample. Then, we used those masses and the spectroscopic metallicities tabulated in \cite{mann15} to compute polynomial empirical relations for \gaia\ $G$, \gbp, \grp, stellar radius, and stellar effective temperature as a function of mass and metallicity. We used the Bayesian Information Criterion to determine the point at which adding additional parameters/changing the functional form no longer produced any significant benefit.

Figure \ref{fig:mdwarfemprels} shows the \citet{mann15,Mann2019} data as a function of mass and colored by metallicity. In each of the diagrams, we do not see any strong trends in either the location or color of the residuals, showing that the mass-metallicity polynomials effectively describe the observational data. Table \ref{tab:emprels} contains the adopted numerical relations. For models within the mass and metallicity range of the \citet{mann15,Mann2019} data, we use each model's mass and metallicity to replace the PARSEC \gbp, \grp, \teff, $R_\star$, \logg, and mean stellar densities with values derived from the empirical polynomial fits displayed in Figure \ref{fig:mdwarfemprels} and quantified in Table \ref{tab:emprels}.

To ensure a smooth transition between the polynomial-modified models and the adjacent PARSEC models, we used a small range of masses and metallicities at which we weight the contribution of the empirical relations for those models. We defined this mass range as 0.63 $<$ $M_\star$ $<$ 0.70 \msun, and the metallicity range as --0.65 $<$ [Fe/H] $<$ --0.50 dex, and computed the difference in the polynomial and evolutionary model predictions. We then scaled this difference based on the product of the mass and metallicity distance from 0.70 \msun\ and --0.65 dex, respectively, and added this difference to the PARSEC model for each parameter of interest. We chose these particular ranges because the \citet{mann15} sample includes stars with metallicities as low as $\approx$--0.6 dex and masses as large as $\approx$0.65 \msun; we also wanted to avoid both (1) a sharp cutoff separating our empirical corrections and the pre-existing models and (2) an extended interpolation over a larger range of masses/metallicities.

Altogether, our modified PARSEC grid includes 2.8 million models, a portion of which is represented in Figure \ref{fig:parsecemp} as an HR diagram. This HR diagram includes a number of notable features:  (1) pre-main sequence models for low-mass stars, which can be seen as the darker colored dots to the right of the main sequence, (2) model aliasing appearing as horizontal ``strips'' of low-mass stars on the main sequence due to computational limitations of the model grid size, (3) models with empirically modified parameters, which appear separated from the evolutionary models for low-mass main sequence stars, and (4) the transition region from empirical relations to evolutionary models.

We used \texttt{isoclassify} \citep{Huber2017,Berger2020a} with the empirically modified PARSEC grid, \gaia\ DR3 \gbp, \grp, parallaxes, positions, and metallicities, and the \texttt{Combined19} map \citep{Drimmel2003,Marshall2006,Green2019} within \texttt{mwdust} \citep{Bovy2016} to correct for dust extinction and derive fundamental stellar parameters. Similar to \citet{Berger2020a}, we also modified \texttt{isoclassify} to utilize uncertainties in \gbp--\grp\ that correspond to a $\sim$3\% uncertainty in \teff\ by fitting a smoothly broken power law (\texttt{astropy}'s \texttt{SmoothlyBrokenPowerLaw1D}) to the model grid's \teff-\gbp--\grp\ relation. We chose the smoothly broken power law because it adequately fits the shape of the \teff-\gbp--\grp\ curve over the range of \teff\ in the model grid with the fewest possible parameters, as compared to the 12th order polynomial fit in \citet{Berger2020a}. We then divided the \teff\ values by the derivative of this relation and multiplied by 3\% to produce a \gbp--\grp\ uncertainty corresponding to 3\% \teff\ uncertainty. Finally, we computed the maximum of the derivative \gbp--\grp\ uncertainty and the measured \gbp\ and \grp\ uncertainties added in quadrature as our adopted color uncertainty. Figure \ref{fig:tefferrfix} shows this comparison and the corresponding power law-derived \teff\ uncertainty. This ensures that we do not drastically underestimate \teff\ uncertainties relative to the 2.0$\pm$0.5\% \teff\ systematic uncertainty floor of interferometry \citep{Tayar2022} given the mmag precision of \gaia\ photometry and the imperfect power law fit to the model grid. For our stellar sample, we find that this procedure gives us a $\approx$2.4\% \teff\ uncertainty for solar-type stars.

\section{Validating the Output Stellar Parameters}\label{sec:valid}

\subsection{Accuracy of Derived Effective Temperatures, Radii, and Masses}\label{sec:teff}

\begin{figure*}
\begin{center}
    \resizebox{0.75\hsize}{!}{\includegraphics{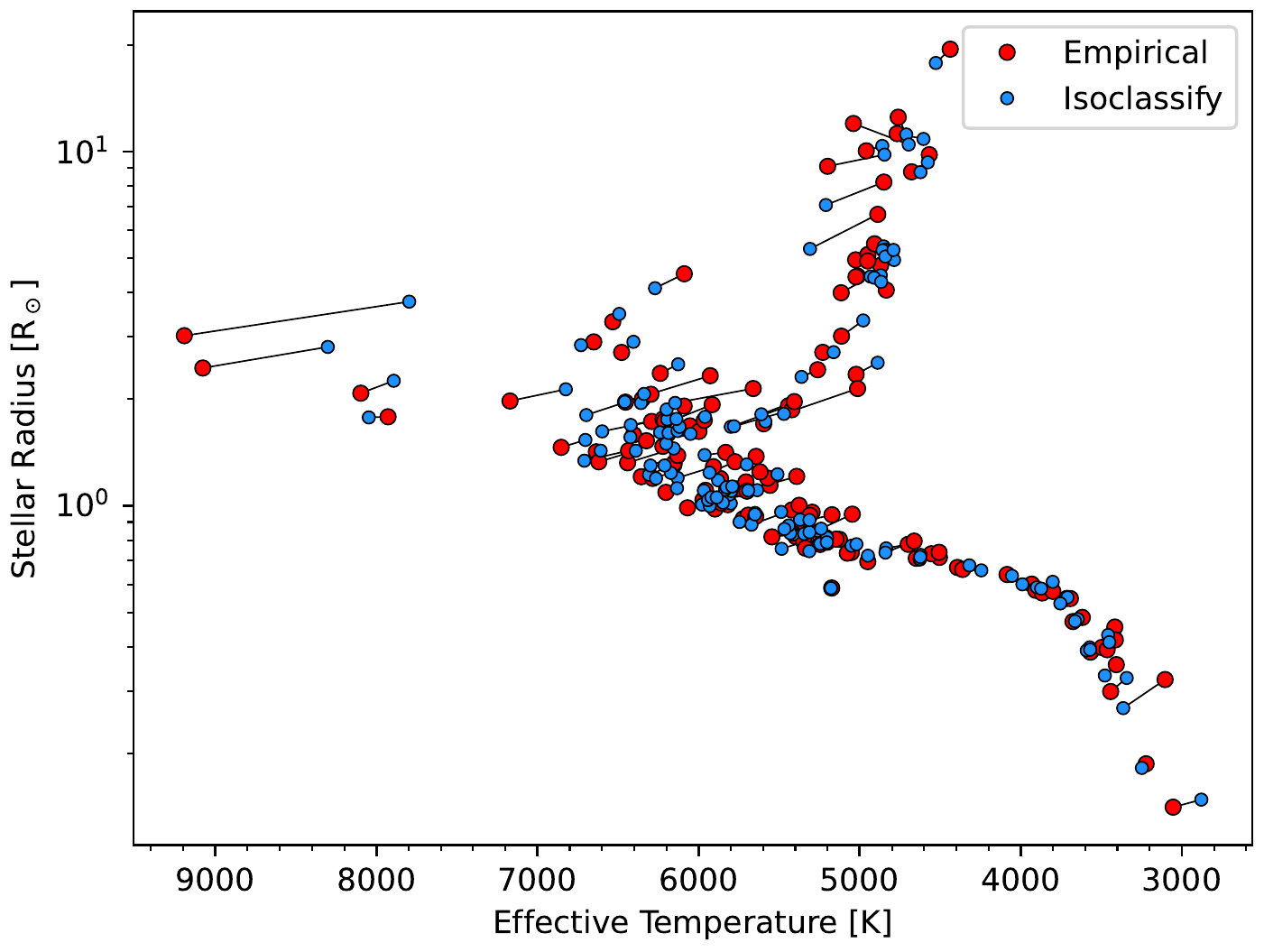}}
\end{center}
\begin{center}
    \resizebox{0.38\hsize}{!}{\includegraphics{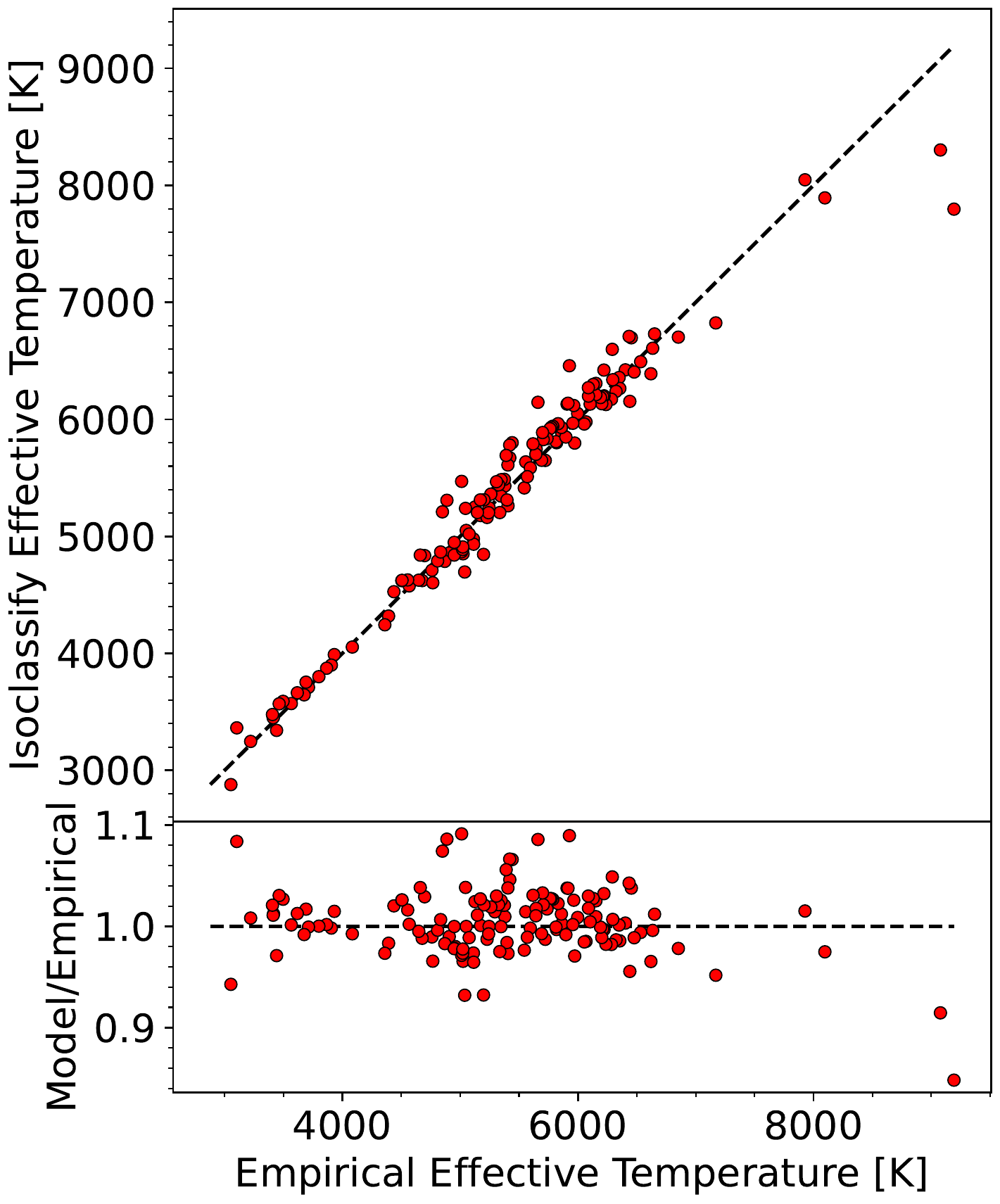}}
    \resizebox{0.375\hsize}{!}{\includegraphics{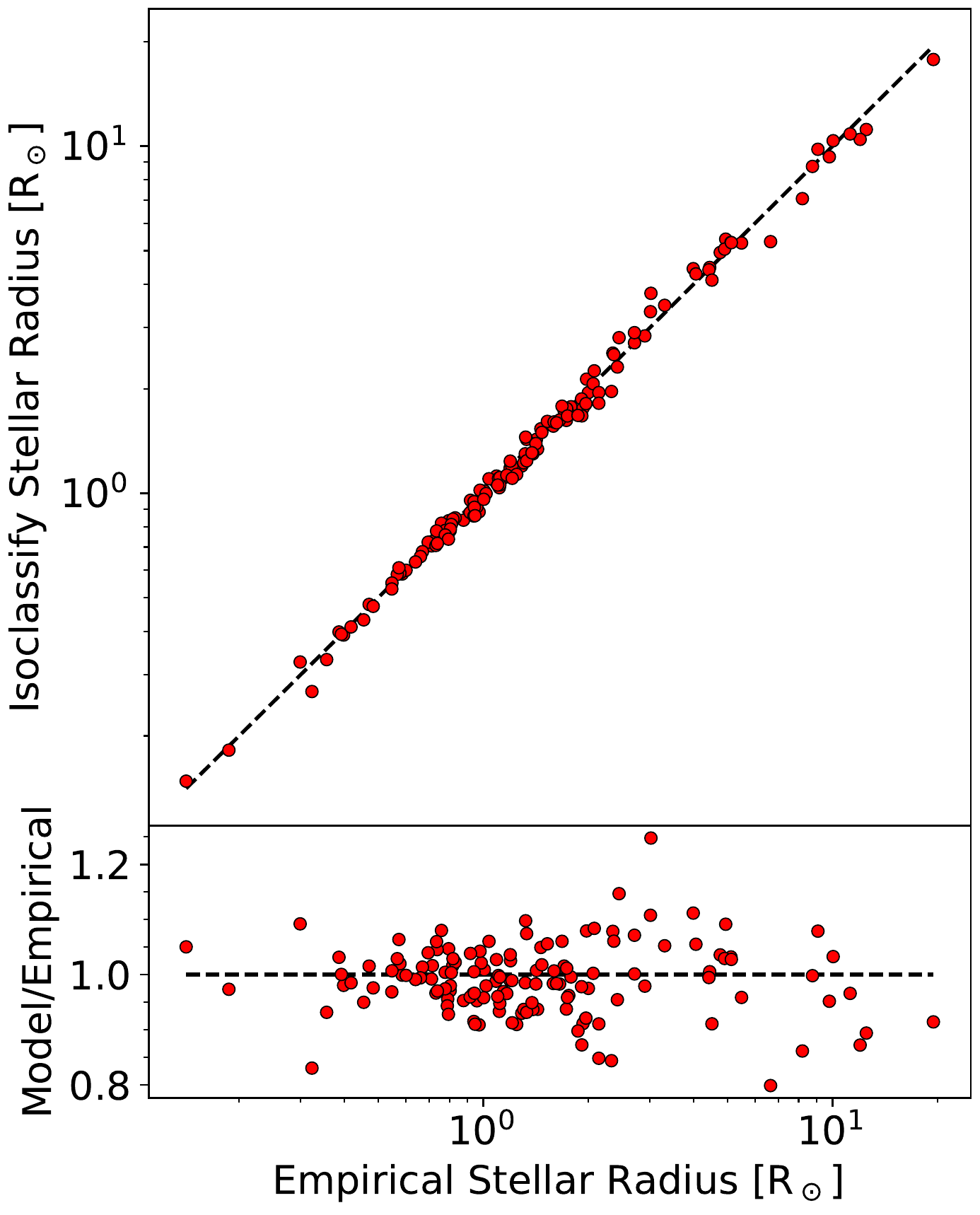}}
\end{center}
\caption{Interferometric \teff\ and radius measurements versus \texttt{isoclassify}-derived results. $Top$: Radius versus \teff\ diagram showing empirical measurements in red and \texttt{isoclassify} results in blue, where black lines connect the separate estimates for the same star. $Left$: Direct comparison of interferometric \teff\ and \texttt{isoclassify} \teff. The black dashed line represents the 1:1 line and the bottom panel shows the corresponding fractional residuals. $Right$: Same as $Left$ but for stellar radius.}
\label{fig:intcomp}
\end{figure*}

\begin{figure*}
\begin{center}
    \resizebox{0.31\hsize}{!}{\includegraphics{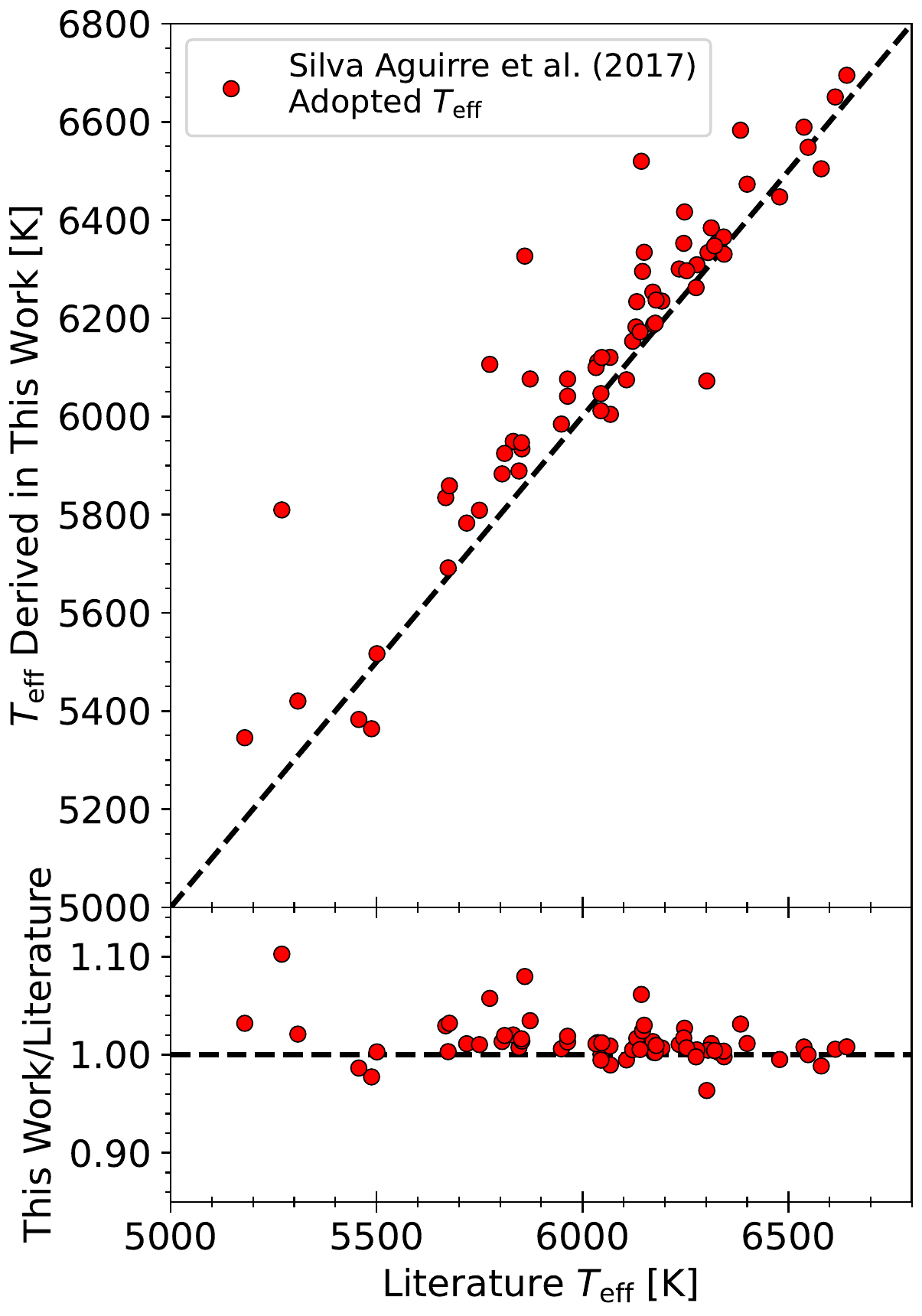}}
    \quad
    \resizebox{0.3\hsize}{!}
    {\includegraphics{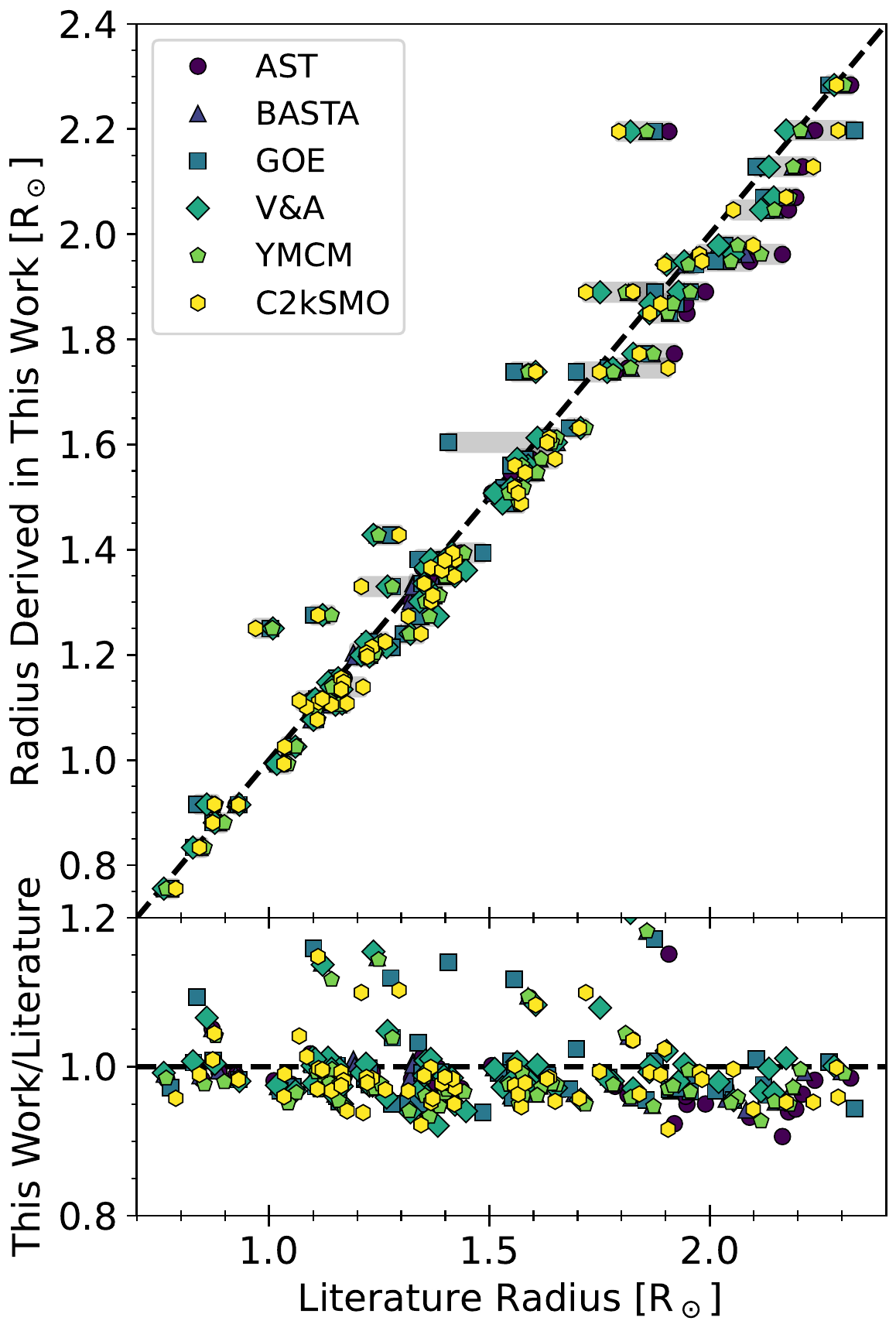}}
    \quad
    \resizebox{0.302\hsize}{!}{\includegraphics{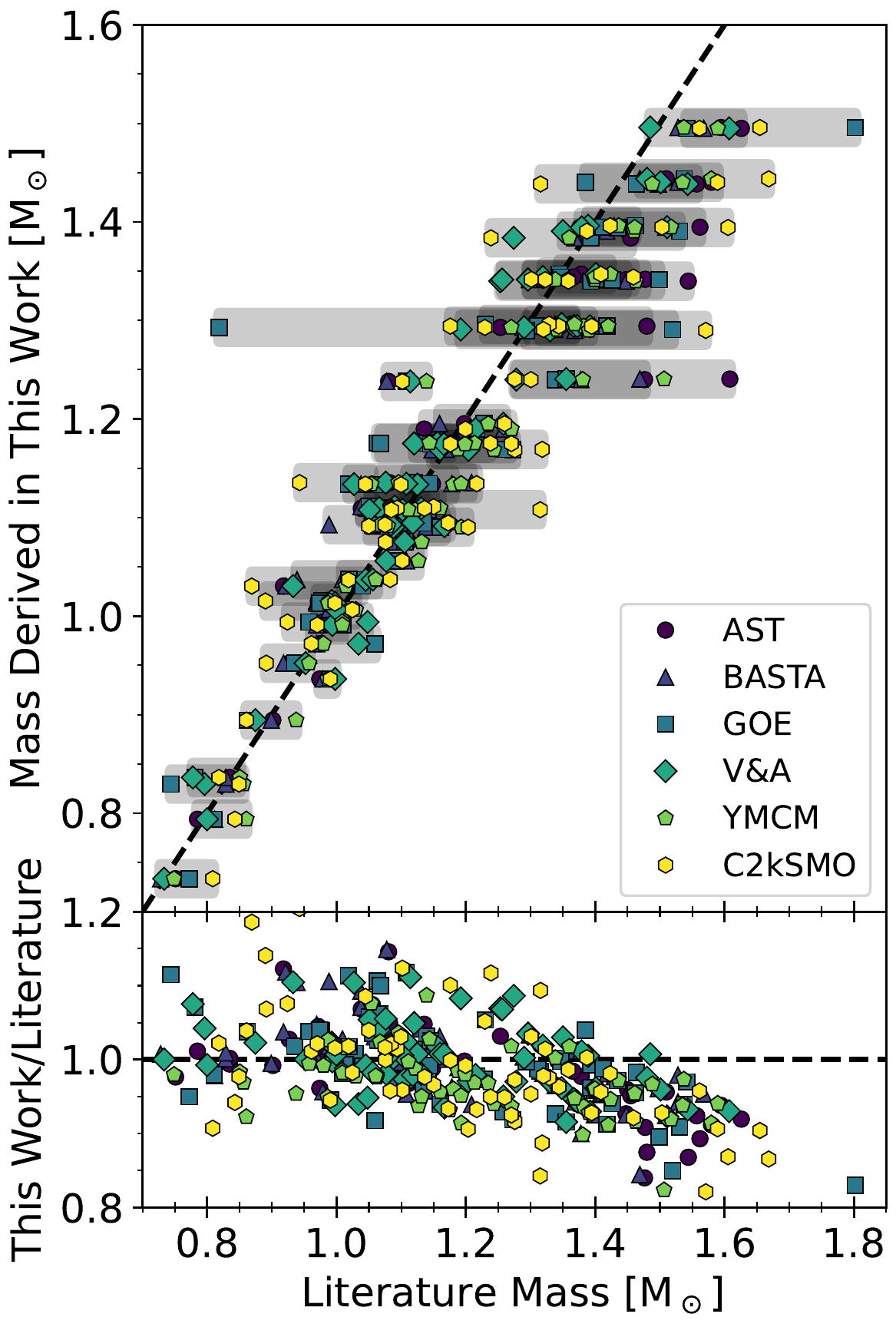}}
\end{center}
\caption{Effective temperatures, radii, and masses derived using our methodology versus those with frequency-modeled asteroseismic properties from a variety of pipelines \citep{silva17}. The black dashed line represents agreement. The various colors/shapes represent the different pipelines. The translucent grey rounded rectangles represent the ranges of parameter estimates from different asteroseismic pipelines for each system, which includes not only seismic differences but also differences between model grids. The bottom panels show the ratio of the parameters derived in this work versus the single or multiple determinations of \citet{silva17}, which produces the curved residual structure apparent for each star in the right two plots.}
\label{fig:legcomp}
\end{figure*}

To ensure our grid-computed stellar effective temperatures and radii are accurate, we compared them to interferometric \teff\ and angular diameter measurements for a sample of 142 stars from \citet{huber12b}, \citet{boyajian13}, \citet{White2018}, \citet{Stokholm2019}, \citet{Rains2020}, and \citet{Karovicova2020,Karovicova2022a,Karovicova2022b}. We followed a similar procedure as above to crossmatch to \gaia\ DR3 sources and correct the photometry and parallaxes for saturation and zeropoint differences, respectively. When we ran \texttt{isoclassify}, we did not correct for extinction given the interferometric sample's proximity to Earth. Of the 142 stars, 118 are within 30 pc, 131 are within 60 pc, and only one (HD 6833) is beyond 200 pc.

Figure \ref{fig:intcomp} shows comparisons of the interferometric stellar parameters and the \texttt{isoclassify}-derived stellar parameters. In general, we show agreement between the models and the interferometric estimates across the entire range of \teff\ and radius, with residual offsets of 0.2\% and 1.0\% and residual scatters of 2.8\% and 5.6\%, compared to the median combined interferometric and \texttt{isoclassify} uncertainties of 2.6\% and 4.6\%, respectively. In addition, we find a systematic underestimation of \teff\ for stars hotter than 9000 K. Since the distance to these stars is less than 30 pc, we can rule out extinction as a cause for this discrepancy. Upon further inspection, the \gaia\ DR3 photometry might be suspect for these objects, as the \gbp--\grp\ colors are much redder than their effective temperature, as the current version\footnote{\url{https://www.pas.rochester.edu/~emamajek/EEM_dwarf_UBVIJHK_colors_Teff.txt}} of the \citet{MamajekTable} table suggests. While the 0.2\% and 1.0\% differences between the scatter and median combined uncertainties are small, they suggest there are additional factors that could produce the excess scatter:  (1) inaccurate photometry/parallaxes, (2) issues with the PARSEC models, and/or (3) issues with the YBC synthetic photometry.

In Figure \ref{fig:legcomp} we compare our derived effective temperatures, radii, and masses to those of the \kep\ LEGACY sample \citep{silva17}, which includes estimates from a number of analysis pipelines for radii and masses. For \teff, radius, and mass, we find offsets of 0.9\%, 1.8\%, and 1.2\% and residual scatters of 1.0\%, 2.4\%, and 5.0\%, with differences between the individual pipelines producing offsets and scatters of similar or smaller size.

\subsection{Accuracy of Derived Stellar Ages}\label{sec:compages}

\subsubsection{Cluster Ages}

\begin{figure}
\resizebox{\hsize}{!}{\includegraphics{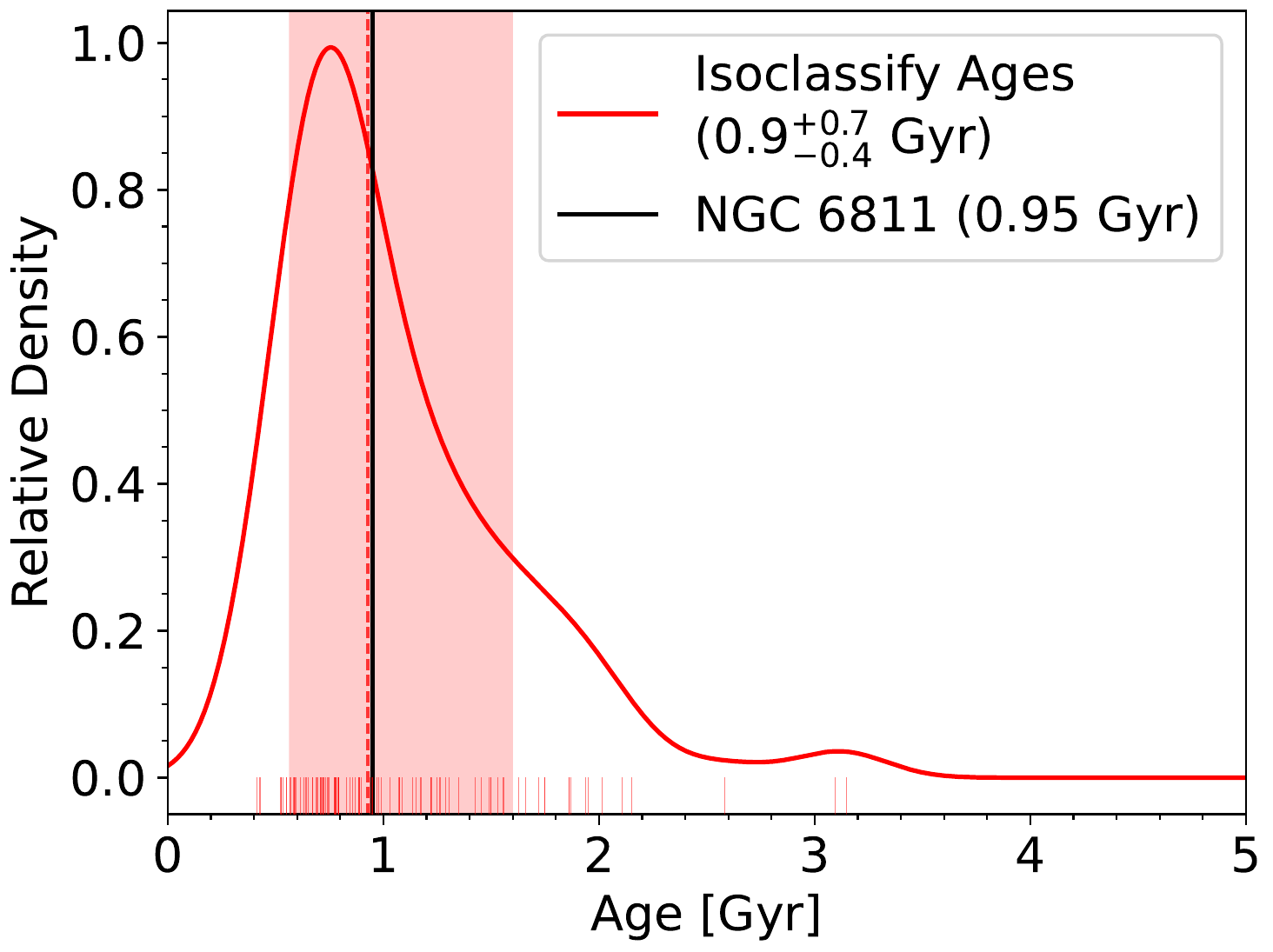}}
\caption{Age comparison for open cluster NGC 6811 at 0.95\,Gyr \citep{Godoy2021}, located within the \kep\ field. The solid red distribution represents Gaussian Kernel Density Estimate (KDE) of the ages of individual stars within each cluster as derived in this work, with the median and the 1$\sigma$ confidence interval represented by the vertical red dashed line and shaded region, respectively. We use Scott's rule \citep{Scott1992} bandwidths to produce the overall distribution. Translucent vertical red lines represent the inferred ages for each star within the sample. The black, solid vertical line represents the cluster ages from the literature in each panel. We only include stars with \teff\ $>$ 6000 K and RUWE $<$ 1.4.}
\label{fig:clusteragecomp}
\end{figure}

To independently confirm our derived stellar ages, we used the NGC 6811 cluster data from \citet{Godoy2021}. We followed a similar procedure as above to crossmatch the NGC 6811 members found by \citet{Godoy2021} to \gaia\ DR3 sources and perform corrections to the parallaxes and photometry, where necessary. We then removed stars not classified as probable members and those without \gaia\ DR3 solutions, leaving us with 338 stars. We used the \citet{Godoy2021} NGC 6811 metallicity ([Fe/H] = 0.03 dex) with 0.15 dex uncertainties for each member, and when we ran \texttt{isoclassify}, we used the \texttt{mwdust} \texttt{allsky} extinction map as above. To ensure that our derived isochrone ages are reliable and minimize the potential for age contaminating binaries \citep{Berger2020a}, we removed all stars with \teff\ $<$ 6000 K and RUWE $>$ 1.4 \citep{Stassun2021}. For the remaining 107 stars, we compared stellar ages.

Figure \ref{fig:clusteragecomp} shows our derived ages for individual stars and their ensemble age versus the \citet{Godoy2021} NGC 6811 cluster age (0.95 Gyr). Our age distribution is in agreement with the cluster's age. Similar to \citet{Berger2020a}, there are a few stars at ages $>$2 Gyr. After plotting the 107 NGC 6811 stars in an HR diagram, we find that the $>$2 Gyr stars are also the coolest and largest subgiants remaining. Although we already removed stars with RUWE $>$ 1.4, some of these members might have very close unresolved companions \citep{Stassun2021} that are less massive and cooler, which can bias our results towards older ages \citep{Berger2020a}.

\subsubsection{Asteroseismic Ages}

\begin{figure}
\resizebox{\hsize}{!}{\includegraphics{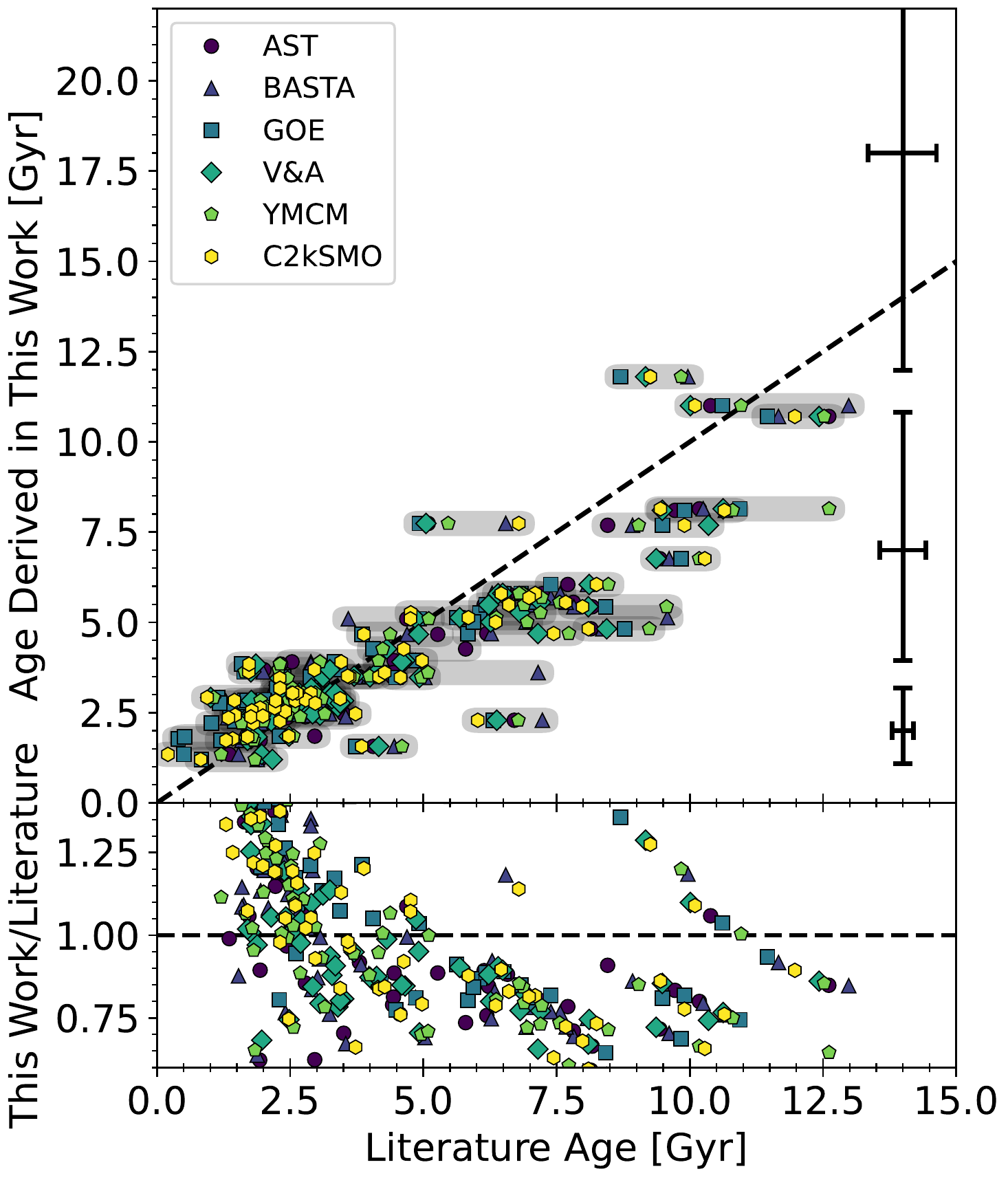}}
\caption{Ages derived using our methodology versus those with frequency-modeled asteroseismic ages from a variety of pipelines \citep{silva17}. The black dashed line represents agreement. The various colors/shapes represent the different pipelines. The translucent grey rounded rectangles represent the ranges of age estimates from different asteroseismic pipelines for each system, which includes not only seismic differences but also differences between model grids. The bottom panel is the ratio of the age derived in this work versus the multiple determinations of \citet{silva17}, which produces the curved residual structure apparent for each star. We have also plotted median error bars in the right-hand portion of the top panel, where, from bottom to top, the error bars represent the median uncertainties of stars with isochrone ages between 0--4 Gyr, 4--8 Gyr, and $>$8 Gyr, respectively.} 
\label{fig:astagecomp}
\end{figure}

We also compared our derived ages to those of \kep\ stars that have asteroseismic ages as in \citet{Berger2020a}. We utilized the ``boutique'' frequency-modeled ages from the \kep\ legacy sample detailed in \cite{silva17}, which includes results from a number of analysis pipelines. In Figure \ref{fig:astagecomp}, the ages that we derive are in reasonable agreement with those provided by a variety of asteroseismic pipelines. The horizontal scatter of the colored points are typically larger than their reported errors, which indicates that systematic pipeline differences dominate. Any deviations from the 1:1 dashed line are sufficiently accounted for by a combination of the typical error bars (bottom right, top panel) and any systematic scatter depending on the asteroseismic pipeline one chooses. The structure in the residuals in the bottom panel results from their representation as a ratio between the age derived in this work and those of \citet{silva17}. In addition, the asteroseismic ages do not fall above the age of the Universe (likely due to a model grid age-cutoff). We also see that at intermediate ages (4--8 Gyr), we may slightly underestimate stellar ages, but our solutions are well-within our reported uncertainties.

Ultimately, we report a median offset and scatter of 0.7\% and 31\% between our isochrone-derived ages and the asteroseismic ages of \citet{silva17}, respectively. We conclude that our isochrone-derived ages are consistent with ages determined through more precise methods within the uncertainties that we report.

\section{Results}\label{sec:results}

\subsection{Stellar Properties}\label{sec:resultsstars}

\begin{figure*}
    \centering
    \resizebox{\hsize}{!}{\includegraphics{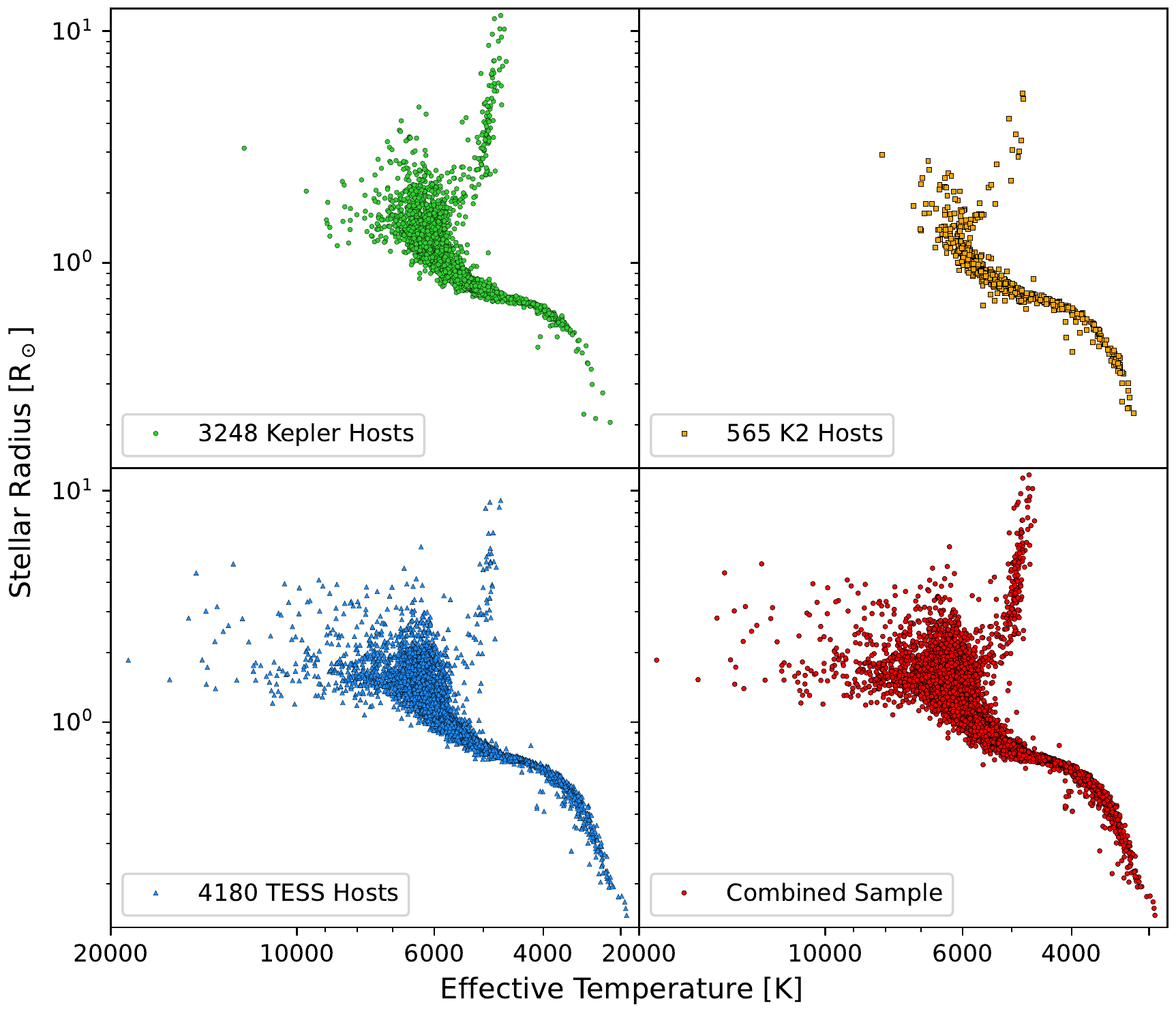}}
    \caption{Stellar radius versus effective temperature for the \kep, \ktwo, \tess, and combined host star samples.}
    \label{fig:HostHR}
\end{figure*}

\begin{figure*}
    \centering
    \resizebox{\hsize}{!}{\includegraphics{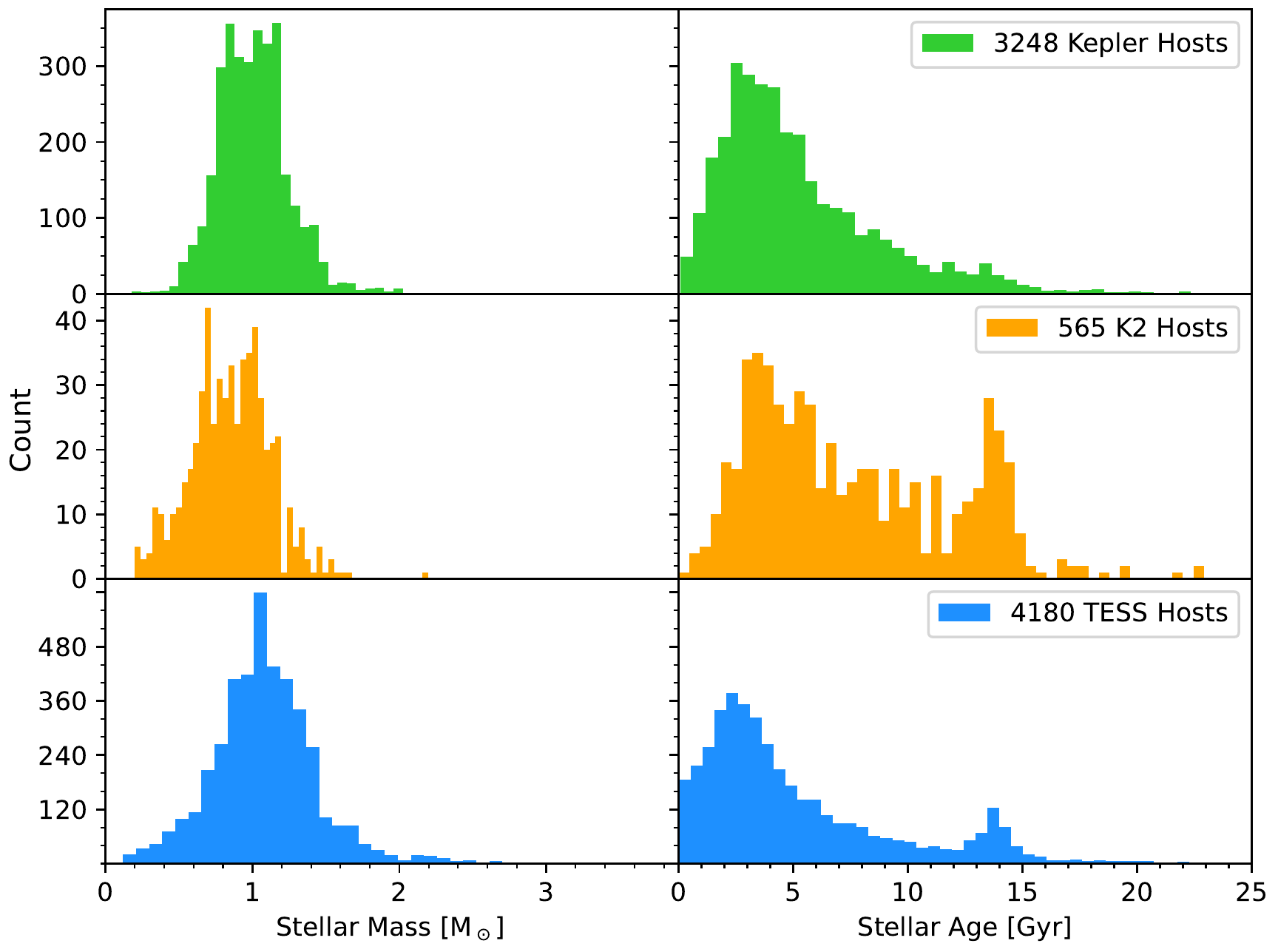}}
    \caption{Histograms of masses and ages for the \kep, \ktwo, and \tess\ host star samples, where the left column contains stellar masses and the right column contains stellar ages. The \kep, \ktwo, and \tess, samples are displayed in each row and ordered, respectively, from top to bottom.}
    \label{fig:HostMassAge}
\end{figure*}

We present our homogeneous stellar properties in Table \ref{tab:stars} and Figure \ref{fig:HostHR}. Figure \ref{fig:HostHR} shows the host star populations broken up into separate panels for each mission and the combined host star sample in the bottom right. Most \kep\ host stars are solar-like stars, as was the priority of the \kep\ mission's target selection \citep{borucki10,batalha10} and as has been confirmed more recently thanks to \gaia\ photometry and parallaxes \citep{Wolniewicz2021}. Of the 3248 \kep\ host stars, only 17 are hotter than 8000 K, while 95 are cooler than 4000 K. Relative to the \ktwo\ and \tess\ host stars, there are more giants with planet candidates in the \kep\ host star sample. This is likely because of \kep's superior sensitivity and baseline when compared to the shorter continuous baselines and increased systematic noise/smaller apertures of \ktwo/\tess.

The upper right panel of Figure \ref{fig:HostHR} displays the \ktwo\ host star sample. There are fewer \ktwo\ hosts than both \kep\ and \tess, mostly because we used the homogeneously derived input sample from \citet{Zink2021}. Due to differences in target selection, most \ktwo\ host stars are lower mass than those observed by \kep. Eighty-nine \ktwo\ stars are cooler than 4000 K and 216 are cooler than 5000 K, as \ktwo\ target selection was community-led and prioritized K and M-dwarfs in its search for planets \citep{dressing13,Howell2014,Dressing2015,Huber2016,Cloutier2020b}. Only one star in our \ktwo\ sample is hotter than 8000 K. In the lower left panel, we plot the \tess\ host star distribution. \tess\ host stars, perhaps unsurprisingly, contain the largest diversity of stellar types, given \tess's all-sky observations. Host stars range in \teff\ from 2934 K up to 18740 K, and of the 4180 hosts, 317 are cooler than 4000 K and 179 are hotter than 8000 K. However, we caution that a large fraction ($\approx$85\%) of the \tess\ host stars are hosts to planet candidates and hence have not been subject to the scrutiny of ground-based follow-up or comprehensive vetting routines.

Together, our \kep+\ktwo+\tess\ host star sample, displayed in the bottom right panel of Figure \ref{fig:HostHR}, represents both the largest and most diverse transiting exoplanet host star sample with homogeneous fundamental parameters yet. This enables a direct comparison of the host stars and their stellar properties and the construction of larger host samples across a wide range of parameter space. For instance, we count 492 host stars with \teff\ $>$ 6000 K and isochrone ages $<$1 Gyr. These stars represent an interesting sample for further study, although we note the potential for many of the planet candidates orbiting these stars, especially those $>$ 10000 K, to be false positives. The ages of exoplanet host stars are particularly useful for determining how exoplanets evolve over time \citep{Soderblom2010,Mann2017,Berger2018,Berger2020b,Berger2022,David2021}.

We plot the distribution of masses and ages in Figure \ref{fig:HostMassAge}. As is clear in Figure \ref{fig:HostHR}, the mass (and age) ranges vary from sample to sample, where \kep\ host stars are mostly close to one solar mass, while \ktwo\ hosts shift to lower masses and \tess\ hosts include a broader distribution of stellar masses. In stellar age, we see that the \kep\ and \ktwo\ distributions are similar except for the peak in stellar age between 13 and 15 Gyr - this is because \ktwo\ focused more heavily on low-mass K and M-dwarfs, which do not produce meaningful isochrone ages and tend to have ages roughly half the maximum age of our model grid (30 Gyr). The \tess\ histogram displays a large number of hosts younger than 3 Gyr, most of which are stars in the high mass tail of the \tess\ mass histogram.

\subsection{Exoplanet Properties}\label{sec:resultsplanets}

\begin{figure*}
    \centering
    \resizebox{\hsize}{!}{\includegraphics{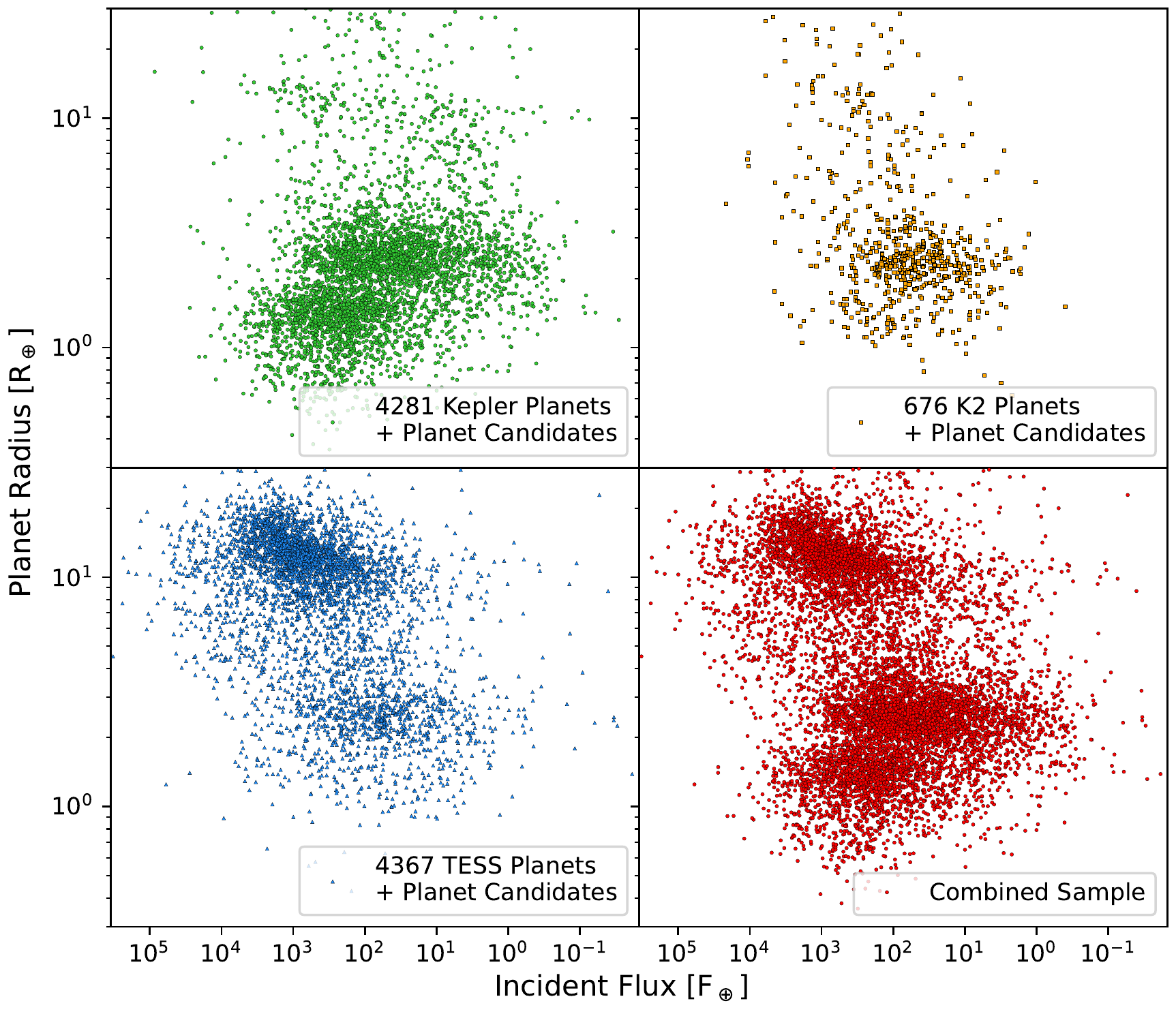}}
    \caption{Planet radius versus incident flux for the \kep, \ktwo, and \tess, and combined planet samples. While most planets are shown, there are 1 \kep\ and 68 \tess\ planets that experience incident fluxes $<$ 0.01 \fearth.}
    \label{fig:PlanetPradIncFlux}
\end{figure*}

\begin{figure}
    \centering
    \resizebox{\hsize}{!}{\includegraphics{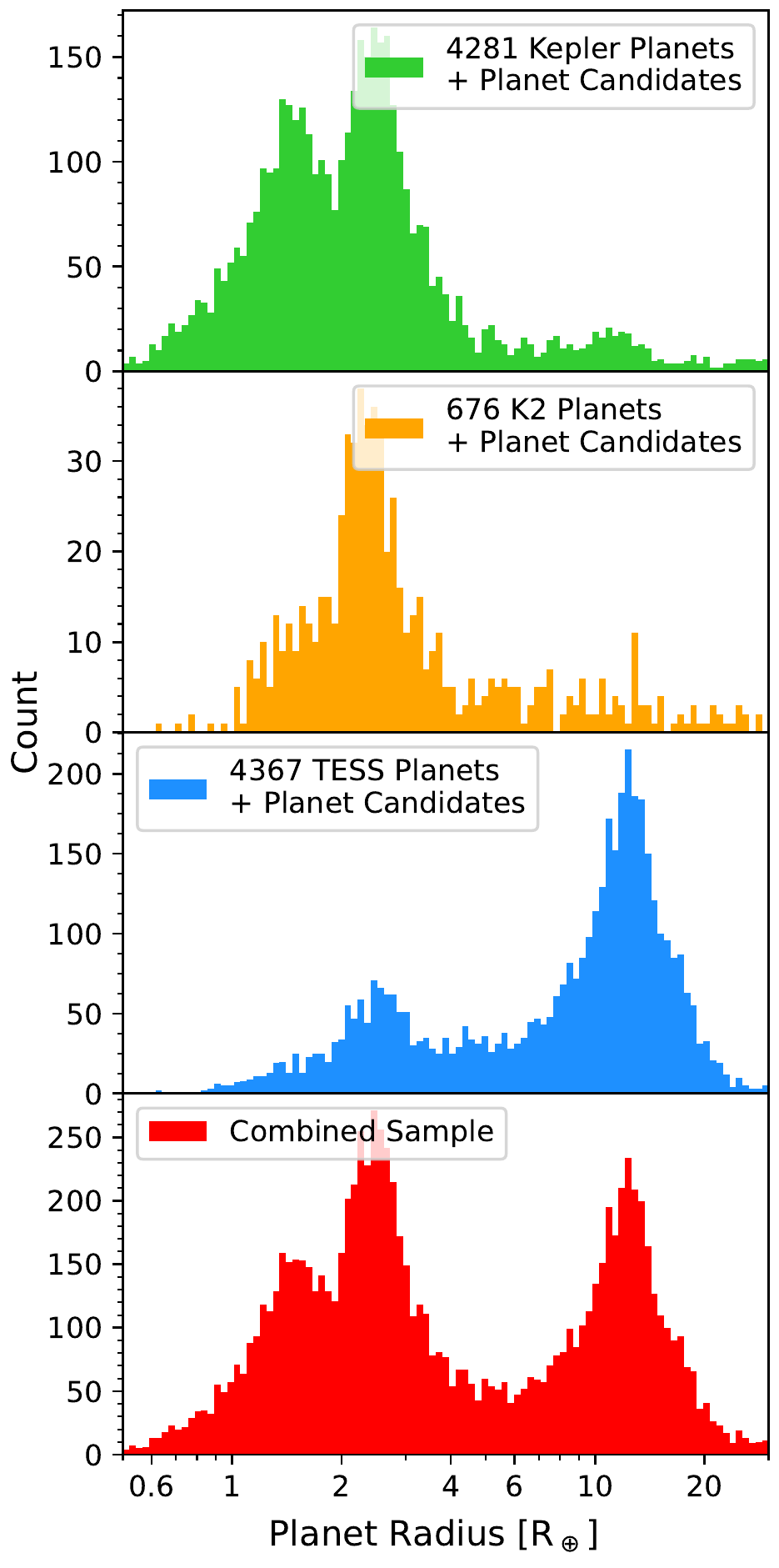}}
    \caption{Planet radius histograms for the \kep, \ktwo, and \tess, and combined planet samples.}
    \label{fig:PradHist}
\end{figure}

As in \citet{Berger2020b}, we utilized the Cumulative KOI table\footnote{accessed 11/5/21} at the NASA Exoplanet Archive \citep{Batalha2013,burke14,rowe15,mullally15,Coughlin2016,Thompson2018,kepcumulative} catalog-provided \rprstar\ and orbital period values, where available, in combination with our stellar parameters and their uncertainties to compute planet radii, semimajor axes, and incident fluxes. Both the Cumulative KOI and \citet{Zink2021} catalogs provide \rprstar\ values for the \kep\ and \ktwo\ planet samples, but the \tess\ Project Candidate table \citep[e.g.][]{Guerrero2021} at the NASA Exoplanet Archive does not. Therefore, we used this \tess\ table's\footnote{accessed 7/27/22} stellar and planet radii to compute \rprstar\ which we then multiplied by our stellar radii to compute \tess\ planet radii. We do not use transit depth measurements to compute \tess\ planet radii because transit depth measurements do not directly translate into planetary radii for planets that have large impact parameters.

We present our planet properties in Table \ref{tab:planets} and Figure \ref{fig:PlanetPradIncFlux}. Similar to the stellar populations in \S\ref{sec:resultsstars}, the planet populations in the planet radius-incident flux diagrams differ from panel-to-panel. In the upper left panel containing \kep\ planets, we can see a few, clear sub-populations of planets. We see two populations of planets smaller than 5 \rearth, separated at by the planet radius gap at $\sim$2 \rearth:  these are the super-Earths ($\lesssim$2 \rearth) and sub-Neptunes ($\gtrsim$2 \rearth and $\lesssim$4 \rearth), as identified by \citet{Fulton2017}. There are many super-Earths due to \kep's sensitivity and baseline, and the sub-Neptune population appears to be well-defined given the drop-off in planets at $>$5 \rearth. Finally, we observe a group of Jupiter-sized planets (10--20 \rearth) at a wide range of incident fluxes.

We show the \ktwo\ planet population in the upper right panel of Figure \ref{fig:PlanetPradIncFlux}. There are far fewer \ktwo\ planets than in either \kep\ or \tess, and while we see the sub-Neptune population, \ktwo\ did not detect many super-Earths. Hence, the planet radius gap is not clear in the \ktwo\ sample due to the small number of super-Earths, although it was detected in \citet{Hardegree2020} and \citet{Zink2021}. The differences in planet radii must come from the utilized stellar radii; we used the same \rprstar\ values as \citet{Zink2021} but used a predominantly photometric isochrone fitting approach as compared to the spectroscopically-dependent random forest classifier of \citet{Hardegree2020}. Comparing the two sets of stellar radii, we find a median offset of $\approx$3\% and typical scatter of $\approx$5\% with some larger outliers. The observation that the \ktwo\ planet radius gap can disappear because of a statistically consistent yet slightly different set of stellar parameters suggests it is not yet robustly determined, unlike the \kep\ planet radius gap.

We plot the \tess\ planets in the bottom left panel, and while there are more \tess\ planets than \kep\ planets, \kep\ has more super-Earths and sub-Neptunes. The vast majority of \tess\ planets are hot Jupiters, which appear to become more inflated at higher incident flux \citep{miller11,Thorngren2016,Grunblatt2016,Grunblatt2017,Grunblatt2019}, although \tess\ has also detected quite a few sub-Neptunes. Due to its shorter observation baselines and reduced sensitivity relative to \kep\ \citep{ricker14}, \tess\ has not detected as many super-Earths. Hence, we do not see a clear gap separating super-Earths and sub-Neptunes. One explanation for the lack of a clear gap in both the \ktwo\ and \tess\ samples are their less-precise \rprstar\ values \citep{Petigura2020}. As above, we caution that a large fraction of the \tess\ planets are planet candidates and upon follow-up some number will be identified as false-positives.

The combined planet sample in the bottom right panel of Figure \ref{fig:PlanetPradIncFlux} represents the largest transiting exoplanet sample with homogeneously derived planet parameters yet. This enables us to increase the number of planets in notable regions of parameter space, such as the hot sub-Neptunian desert \citep[$R_{\mathrm{p}}$ = 2.2--3.8 \rearth\ and $F_{\mathrm{p}}$ $>$ 650 \fearth,][]{Lundkvist2016,Berger2018c,Berger2020b}, the habitable zone \citep[$F_{\mathrm{p}}$ = 0.25--1.50 \fearth,][]{Kane2016,Berger2018c,Berger2020b}, and the planet radius gap \citep{Fulton2017,Fulton2018,VanEylen2018,Berger2018c,Berger2020b,Petigura2020,Petigura2022}. Moreover, the populations of inflated Jupiters, both hot and warm/cold \citep{Berger2018c,Berger2020b}, warrant additional scrutiny.

We take a closer look at the \kep, \ktwo, \tess, and combined planet radius histograms in Figure \ref{fig:PradHist}. Similar to our conclusions based on the various panels in Figure \ref{fig:PlanetPradIncFlux}, we see no clear gap in either the \ktwo\ or \tess\ planet radius histograms. We emphasize that these are raw numbers and are not completeness-corrected. Comparing the \kep\ and combined samples, it is clear that \kep\ provides superior gap resolution, and the addition of \ktwo\ and \tess\ planets makes it less-clear.

\section{Conclusion}\label{sec:conclusion}

We presented the \gaia-\kep-\tess-Host Stellar Properties Catalog and the corresponding catalog of homogeneous exoplanet properties, which contain \nstars\ stars and \nplanets\ planets, respectively. We used \texttt{isoclassify} and \gaia\ DR3 to compute precise, homogeneous \teff, \logg, masses, radii, mean stellar densities, luminosities, ages, distances, and V-band extinctions for 3248, 565, and 4180 \kep, \ktwo, and \tess\ stars, respectively. We compared our stellar properties to studies using fundamental and precise constraints, such as interferometry and asteroseismology, and find good agreement between our \teff, radii, and ages and those in the literature. In addition, we provide planet radii, semimajor axes, and incident fluxes for 4281, 676, and 4367 \kep, \ktwo, and \tess\ planets, respectively, and find that the exoplanet radius gap is less prominent in \ktwo\ and \tess\ and combined samples than it is in the \kep\ sample alone. We provide our stellar and planet catalogs as machine-readable tables associated with this article and as a High-Level Science Product (HLSP) at the Mikulski Archive for Space Telescopes (MAST). Finally, we identify over 1000 hot Jupiter planet candidates, 150 planets within the hot sub-Neptunian desert, and more than 400 young host stars as potential opportunities for testing the theories of planet formation and evolution.

\software{\texttt{Astropy} \citep{astropy:2013, astropy:2018, astropy:2022}, \texttt{emcee} \citep{Foreman-Mackey2013}, \texttt{GNU Parallel} \citep{GNUparallel}, \texttt{kiauhoku} \citep{Claytor2020}, \texttt{matplotlib} \citep{Matplotlib}, \texttt{numpy} \citep{numpy}, \texttt{pandas} \citep{Pandas}, \texttt{scipy} \citep{Scipy}, \texttt{TOPCAT} \citep{topcat}}

\facility{Exoplanet Archive}

\begin{acknowledgments}
We thank Yang Chen for providing access to synthetic photometry for the custom-interpolated PARSEC model grid. This research has made use of the NASA Exoplanet Archive, which is operated by the California Institute of Technology, under contract with the National Aeronautics and Space Administration under the Exoplanet Exploration Program. We acknowledge the use of public TOI Release data from pipelines at the TESS Science Office and at the TESS Science Processing Operations Center. This work has made use of data from the European Space Agency (ESA) mission {\it Gaia} (\url{https://www.cosmos.esa.int/gaia}), processed by the {\it Gaia} Data Processing and Analysis Consortium (DPAC, \url{https://www.cosmos.esa.int/web/gaia/dpac/consortium}). Funding for the DPAC has been provided by national institutions, in particular the institutions
participating in the {\it Gaia} Multilateral Agreement. T.A.B.’s research was supported by an appointment to the NASA Postdoctoral Program at the NASA Goddard Space Flight Center, administered by Universities Space Research Association and Oak Ridge Associated Universities under contract with NASA. D.H. acknowledges support from the Alfred P. Sloan Foundation and the National Aeronautics and Space Administration (80NSSC19K0597, 80NSSC21K0652).
\end{acknowledgments}

\begin{deluxetable*}{lcccccccccr}
\tabletypesize{\scriptsize}
\tablewidth{0pt}
\tablecolumns{11}
\tablecaption{\gaia-\kep-\tess-Host Stellar Output Parameters}
\tablehead{\colhead{Star ID} & \colhead{\teff\ [K]} & \colhead{\logg\ [dex]} & \colhead{[Fe/H]} & \colhead{$M_\star$ [$\mathrm{M_\odot}$]} & \colhead{$R_\star$ [$\mathrm{R_\odot}$]} & \colhead{$\rho_\star$ [$\mathrm{\rho_\odot}$]} & \colhead{$L_\star$ [$\mathrm{L_\odot}$]} & \colhead{Age [Gyr]} & \colhead{Distance [pc]} & \colhead{$A_V$ [mag]}}
\startdata
kic10858832 & 5812$^{+106}_{-124}$ & 4.424$^{+0.035}_{-0.058}$ & 0.140$^{+0.131}_{-0.138}$ & 1.054$^{+0.042}_{-0.061}$ & 1.040$^{+0.041}_{-0.030}$ & 0.927$^{+0.103}_{-0.146}$ & 1.122$^{+0.048}_{-0.045}$ & 2.86$^{+3.75}_{-2.17}$ & 853.5$^{+15.4}_{-15.0}$ & 0.110 \\
kic2571238 & 5444$^{+99}_{-110}$ & 4.510$^{+0.033}_{-0.058}$ & 0.055$^{+0.139}_{-0.135}$ & 0.915$^{+0.043}_{-0.056}$ & 0.880$^{+0.035}_{-0.026}$ & 1.334$^{+0.142}_{-0.211}$ & 0.620$^{+0.019}_{-0.019}$ & 4.29$^{+6.25}_{-3.33}$ & 218.9$^{+3.1}_{-3.0}$ & 0.000 \\
kic8628665 & 5895$^{+113}_{-122}$ & 4.450$^{+0.030}_{-0.051}$ & -0.021$^{+0.135}_{-0.131}$ & 1.034$^{+0.038}_{-0.059}$ & 1.000$^{+0.037}_{-0.030}$ & 1.024$^{+0.103}_{-0.144}$ & 1.093$^{+0.062}_{-0.058}$ & 2.31$^{+3.43}_{-1.79}$ & 1079.5$^{+26.2}_{-25.5}$ & 0.275 \\
kic10328393 & 4818$^{+84}_{-106}$ & 4.596$^{+0.033}_{-0.058}$ & -0.092$^{+0.140}_{-0.135}$ & 0.735$^{+0.041}_{-0.039}$ & 0.715$^{+0.024}_{-0.023}$ & 2.001$^{+0.221}_{-0.304}$ & 0.250$^{+0.008}_{-0.010}$ & *7.98$^{+11.15}_{-6.15}$ & 346.7$^{+5.1}_{-5.0}$ & 0.027 \\
epic210577548 & 5381$^{+114}_{-121}$ & 4.531$^{+0.038}_{-0.066}$ & -0.098$^{+0.136}_{-0.135}$ & 0.854$^{+0.040}_{-0.059}$ & 0.829$^{+0.037}_{-0.029}$ & 1.486$^{+0.186}_{-0.260}$ & 0.525$^{+0.017}_{-0.016}$ & 5.61$^{+8.01}_{-4.34}$ & 284.3$^{+4.3}_{-4.1}$ & 0.604 \\
epic220294712 & 6294$^{+135}_{-153}$ & 4.360$^{+0.037}_{-0.056}$ & -0.127$^{+0.121}_{-0.144}$ & 1.132$^{+0.044}_{-0.060}$ & 1.159$^{+0.047}_{-0.035}$ & 0.718$^{+0.085}_{-0.111}$ & 1.915$^{+0.085}_{-0.080}$ & 1.88$^{+2.31}_{-1.37}$ & 428.0$^{+6.6}_{-6.4}$ & 0.192 \\
epic211711685 & 5484$^{+95}_{-106}$ & 4.496$^{+0.033}_{-0.059}$ & 0.111$^{+0.132}_{-0.138}$ & 0.936$^{+0.040}_{-0.044}$ & 0.906$^{+0.036}_{-0.026}$ & 1.256$^{+0.132}_{-0.199}$ & 0.676$^{+0.022}_{-0.021}$ & 4.00$^{+5.78}_{-3.09}$ & 288.7$^{+4.2}_{-4.1}$ & 0.000 \\
epic251584580 & 4830$^{+133}_{-242}$ & 4.601$^{+0.044}_{-0.064}$ & -0.352$^{+0.128}_{-0.123}$ & 0.696$^{+0.021}_{-0.042}$ & 0.686$^{+0.028}_{-0.025}$ & 2.103$^{+0.304}_{-0.355}$ & 0.233$^{+0.014}_{-0.030}$ & *9.59$^{+11.13}_{-7.18}$ & 379.8$^{+6.2}_{-5.9}$ & 0.137 \\
tic429501231 & 5779$^{+164}_{-188}$ & 4.371$^{+0.073}_{-0.082}$ & -0.186$^{+0.134}_{-0.150}$ & 0.914$^{+0.080}_{-0.081}$ & 1.025$^{+0.060}_{-0.049}$ & 0.828$^{+0.199}_{-0.178}$ & 1.073$^{+0.055}_{-0.051}$ & 8.01$^{+6.05}_{-5.06}$ & 527.8$^{+11.3}_{-11.1}$ & 0.110 \\
tic255685030 & 5519$^{+91}_{-98}$ & 4.502$^{+0.028}_{-0.050}$ & 0.120$^{+0.131}_{-0.138}$ & 0.957$^{+0.037}_{-0.045}$ & 0.909$^{+0.031}_{-0.024}$ & 1.270$^{+0.114}_{-0.174}$ & 0.696$^{+0.023}_{-0.021}$ & 3.16$^{+4.84}_{-2.47}$ & 157.5$^{+2.2}_{-2.2}$ & 0.053 \\
tic238624131 & 5924$^{+115}_{-141}$ & 4.384$^{+0.041}_{-0.062}$ & 0.148$^{+0.130}_{-0.143}$ & 1.096$^{+0.057}_{-0.062}$ & 1.110$^{+0.048}_{-0.033}$ & 0.792$^{+0.101}_{-0.133}$ & 1.376$^{+0.053}_{-0.049}$ & 2.84$^{+3.39}_{-2.12}$ & 330.6$^{+4.9}_{-4.8}$ & 0.247 \\
tic391903064 & 5721$^{+107}_{-126}$ & 4.451$^{+0.035}_{-0.059}$ & 0.088$^{+0.140}_{-0.133}$ & 1.012$^{+0.044}_{-0.060}$ & 0.987$^{+0.039}_{-0.028}$ & 1.038$^{+0.117}_{-0.163}$ & 0.950$^{+0.034}_{-0.032}$ & 3.30$^{+4.48}_{-2.51}$ & 78.8$^{+1.1}_{-1.1}$ & 0.089 \\
\enddata
\tablecomments{KIC ID, effective temperature, surface gravity, surface metallicity, stellar mass, stellar radius, density, luminosity, age, distance, and $V$-magnitude extinction, output from our isochrone fitting routine detailed in \S\ref{sec:methods}. Asterisks are appended to stellar ages that are uninformative (\teff\ $<$ 5000 K). A subset of our output parameters is provided here to illustrate the form and format. The full table, in machine-readable format, can be found online.} \label{tab:stars}
\end{deluxetable*}

\begin{deluxetable*}{lccrrrrr}
\tabletypesize{\scriptsize}
\tablewidth{0pt}
\tablecolumns{8}
\tablecaption{\gaia-\kep-\tess\ Planet Parameters}
\tablehead{
\colhead{Star ID} & \colhead{Planet ID} & \colhead{Disposition} & \colhead{$P$ [days]} & \colhead{$R_{\mathrm{p}}/R_\star$} & \colhead{$R_{\mathrm{p}}$ [\rearth]} & \colhead{$a$ [AU]} & \colhead{$F_{\mathrm{p}}$ [\fearth]}}
\def\arraystretch{1.0}
\startdata
kic10797460 & K00752.01 & CP & 9.488036$^{+0.000028}_{-0.000028}$ & 0.02234$^{+0.00083}_{-0.00053}$ & 2.20$^{+0.12}_{-0.09}$ & 0.08577$^{+0.00119}_{-0.00182}$ & 99.190$^{+8.149}_{-7.296}$ \\
kic10797460 & K00752.02 & CP & 54.418383$^{+0.000248}_{-0.000248}$ & 0.02795$^{+0.00908}_{-0.00135}$ & 2.76$^{+0.90}_{-0.16}$ & 0.27482$^{+0.00381}_{-0.00584}$ & 9.661$^{+0.794}_{-0.711}$ \\
kic10811496 & K00753.01 & PC & 19.899140$^{+0.000015}_{-0.000015}$ & 0.15405$^{+5.03429}_{-0.04218}$ & 18.82$^{+615.00}_{-5.23}$ & 0.14628$^{+0.00290}_{-0.00369}$ & 61.417$^{+5.571}_{-5.550}$ \\
kic10854555 & K00755.01 & CP & 2.525592$^{+0.000004}_{-0.000004}$ & 0.02406$^{+0.00375}_{-0.00152}$ & 2.42$^{+0.40}_{-0.18}$ & 0.03550$^{+0.00053}_{-0.00080}$ & 629.842$^{+46.532}_{-48.596}$ \\
epic201155177 & epic201155177.01 & KP & 6.689040$^{+0.000796}_{-0.000871}$ & 0.02850$^{+0.00100}_{-0.00100}$ & 2.15$^{+0.09}_{-0.10}$ & 0.06105$^{+0.00109}_{-0.00123}$ & 49.117$^{+3.183}_{-4.935}$ \\
epic201208431 & epic201208431.01 & KP & 10.004208$^{+0.000881}_{-0.000858}$ & 0.03520$^{+0.00100}_{-0.00100}$ & 2.47$^{+0.09}_{-0.09}$ & 0.07737$^{+0.00086}_{-0.00086}$ & 17.275$^{+1.036}_{-0.857}$ \\
epic201247497 & epic201247497.01 & PC & 2.754012$^{+0.000122}_{-0.000132}$ & 0.09170$^{+0.02800}_{-0.01900}$ & 6.82$^{+2.09}_{-1.43}$ & 0.03339$^{+0.00040}_{-0.00066}$ & 116.013$^{+15.740}_{-14.916}$ \\
epic201295312 & epic201295312.01 & KP & 5.656766$^{+0.000276}_{-0.000323}$ & 0.01710$^{+0.00000}_{-0.00000}$ & 2.73$^{+0.14}_{-0.13}$ & 0.06523$^{+0.00160}_{-0.00146}$ & 567.312$^{+36.131}_{-32.633}$ \\
tic88863718 & toi1001.01 & PC & 1.931671$^{+0.000008}_{-0.000008}$ & 0.04701$^{+0.01480}_{-0.01480}$ & 10.05$^{+3.20}_{-3.20}$ & 0.03351$^{+0.00039}_{-0.00045}$ & 7303.777$^{+360.938}_{-354.659}$ \\
tic65212867 & toi1007.01 & PC & 6.998921$^{+0.000014}_{-0.000014}$ & 0.05012$^{+0.00354}_{-0.00354}$ & 14.40$^{+1.22}_{-1.22}$ & 0.08440$^{+0.00097}_{-0.00158}$ & 1506.786$^{+71.140}_{-80.125}$ \\
tic231663901 & toi101.01 & KP & 1.430369$^{+0.000001}_{-0.000001}$ & 0.13623$^{+0.00954}_{-0.00954}$ & 13.23$^{+1.04}_{-1.00}$ & 0.02429$^{+0.00031}_{-0.00054}$ & 1301.559$^{+59.105}_{-73.562}$ \\
tic114018671 & toi1011.01 & PC & 2.469696$^{+0.000579}_{-0.000579}$ & 0.01415$^{+0.01856}_{-0.01856}$ & 1.38$^{+1.82}_{-1.82}$ & 0.03468$^{+0.00055}_{-0.00099}$ & 547.740$^{+24.947}_{-35.398}$ \\
\enddata
\tablecomments{Host star ID, planet ID, disposition, orbital period, \rprstar, planet radius, semimajor axis, and incident flux values and their uncertainties for all planets contained within our catalog. Dispositions are designated as KP, CP, PC, APC, or LPPC for Known Planet, Confirmed Planet, Planetary Candidate, Ambiguous Planetary Candidate, or Long Period Planetary Candidate, respectively. A subset of our planet parameters is provided here to illustrate the form and format. The full table, in machine-readable format, can be found online.} \label{tab:planets}
\end{deluxetable*}

\bibliography{references}{}
\bibliographystyle{aasjournal}


\end{document}